\documentclass{article}

\usepackage{PRIMEarxiv}

\usepackage[utf8]{inputenc} 
\usepackage[T1]{fontenc}    
\usepackage{hyperref}       
\usepackage{url}            
\usepackage{booktabs}       
\usepackage{amsfonts}       
\usepackage{nicefrac}       
\usepackage{microtype}      
\usepackage{lipsum}
\usepackage{fancyhdr}       
\usepackage{graphicx}       
\graphicspath{{media/}}     
\usepackage{subfig}
\usepackage{enumitem}
\setlist[itemize]{align=parleft,left=0pt..1em}

\AtBeginDocument{%
  \providecommand\BibTeX{{%
    \normalfont B\kern-0.5em{\scshape i\kern-0.25em b}\kern-0.8em\TeX}}}

\title{A Fresh Perspective on DNN Accelerators by Performing Holistic Analysis Across Paradigms
}

\author{
Tom Glint\\
IIT Gandhinagar \\
  \texttt{tom.issac@iitgn.ac.in} \\
  \and
  Chandan Kumar Jha\\
  DFKI Bremen \\
  \texttt{chandan.jha@dfki.de} \\
  \and
  Manu Awasthi\\
  Ashoka University \\
  \texttt{manu.awasthi@ashoka.edu.in} \\
  \and
  Joycee Mekie\\
  IIT Gandhinagar \\
  \texttt{joycee@iitgn.ac.in} \\
}

\begin{document}
\maketitle

\begin{abstract}
Traditional computers with von Neumann architecture are unable to meet the latency and scalability challenges of Deep Neural Network (DNN) workloads. Various DNN accelerators based on Conventional compute Hardware Accelerator (CHA), Near-Data-Processing (NDP) and Processing-in-Memory (PIM) paradigms have been proposed to meet these challenges. Our goal in this work is to perform a rigorous comparison among the state-of-the-art accelerators from DNN accelerator paradigms, we have used unique layers from MobileNet, ResNet, BERT, and DLRM of MLPerf Inference benchmark for our analysis. 
The detailed models are based on hardware-realized state-of-the art designs. We observe that for memory-intensive Fully Connected Layer (FCL) DNNs, NDP based accelerator is 10.6$\times$ faster than the state-of-the-art CHA and 39.9$\times$ faster than PIM based accelerator for inferencing. For compute-intensive image classification and object detection DNNs, the state-of-the-art CHA is $\sim$10$\times$ faster than NDP and $\sim$2000$\times$ faster than the PIM-based accelerator for inferencing. PIM-based accelerators are suitable for DNN applications where energy is a constraint ($\sim$2.7$\times$ and $\sim$21$\times$ lower energy for CNN and FCL applications, respectively, than conventional ASIC systems). Further, we identify architectural changes (such as increasing memory bandwidth, buffer reorganization) that can increase throughput (up to linear increase) and lower energy (up to linear decrease) for ML applications with a detailed sensitivity analysis of relevant components in CHA, NDP and PIM based accelerators. 
\end{abstract}


\section{Introduction}\label{Introduction}
\begin{figure*}[htpb]
    \centering

    \includegraphics[width=\linewidth]{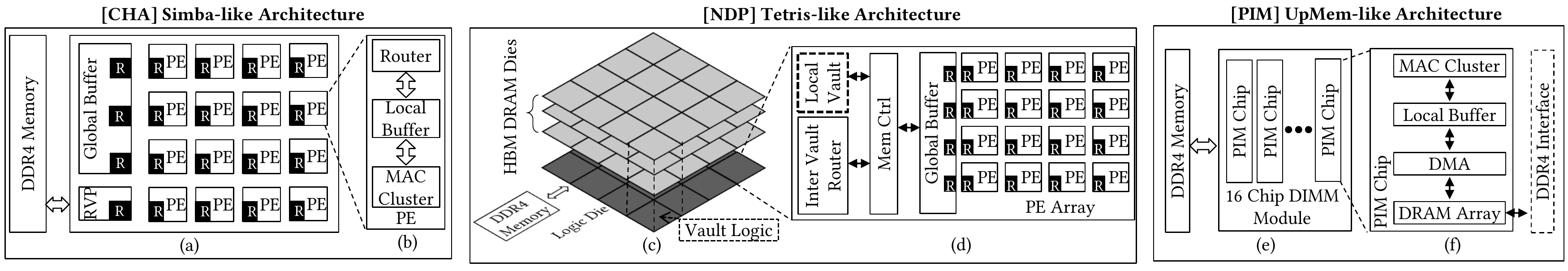}
  
    \caption{(a) Simba-like architecture: The ASIC Chip with Global buffer and an array of $4\times4$ PEs. (b) A single PE of Simba-like architecture. (c) Tetris-like architecture: The logic die and 3D DRAM memory with 16 vaults. (d) Compute logic associated with a single vault, with 16 PEs in $4\times4$ array configuration. (e) UpMem-like architecture: 16 chip DIMM module connected to DDR4 interface. (f) PIM chip with a MAC cluster containing 8 MAC units.}

    \label{fig:Architectures}
    \end{figure*}

Deep Neural Networks (DNN) have evolved to do image classification, object detection, language processing, and recommendation tasks in mobile, edge, and datacenter applications \cite{mattson2019mlperf}. 
These applications have specific latency and energy constraints based on the \emph{scenarios} where these applications are deployed \cite{parashar2019timeloop}. 
Since traditional computers with von Neumann architecture are not suitable for meeting these constraints, various DNN accelerator architecture paradigms have been proposed \cite{surveyofhardware,UpMem,tetris,Simba2,damov,gomez2021benchmarking}.
As shown in Fig.~\ref{fig:Architectures}, they can be broadly classified into Conventional Hardware Accelerators (CHA) \cite{Simba2,damov}, Near Data Processors (NDP) \cite{tetris, damov}, and Processing in Memory (PIM) Accelerators \cite{UpMem,gomez2021benchmarking}. 
Application-specific integrated circuit (ASIC) based CHA DNN accelerators are designed to provide fast compute capability with the help of operation specific circuits but are limited by DRAM interface bandwidth due to limited pin out count~\cite{damov}. 
Simba~\cite{Simba2}, as shown in Fig.~\ref{fig:Architectures}a is a hardware realized example of CHA.
An alternative to CHA, PIM-based accelerators are proposed to overcome the memory bandwidth bottleneck. 
Commercial PIM based accelerators use 2D DRAM process to fabricate compute elements on the same chip as memory. 
However, they are limited to simple core designs without dedicated single cycle hardware multipliers due to the DRAM fabrication process \cite{damov, gomez2021benchmarking}. Fig.~\ref{fig:Architectures}e shows UpMem \cite{UpMem}, a state-of-the-art PIM based accelerator.
NDP-based DNN accelerator architectures, which reduce data movement costs by placing compute capability close to memory, are a middle ground between the other two architectures \cite{damov}. 
NDP accelerators based on 3D-stacked memory integrate DRAM layers with a logic layer which consists of processing elements.
Still, the logic layer has area and thermal constraints, limiting the capabilities of processing components. 
Tetris \cite{tetris}, shown in Fig.~\ref{fig:Architectures}c, is such an NDP architecture. While such a varying range of architectures exist, there is no works which performs in-depth comparisons among these three architectural paradigms.
Traditionally the existing works have limit the comparison of DNN accelerators against CPUs and GPUs. 

To the best of our knowledge this is the first work where in depth analysis and comparison has been done among these three paradigms. We identify the DNN architectural paradigm suitable for different ML workloads by detailed architectural analysis, for different \emph{scenarios} where latency, throughput, or energy efficiency are essential. Further, we identify changes that can be made to the components of each architectural paradigm to improve latency and energy efficiency for ML workloads by carrying out sensitivity analysis. We also identify the upper limits of performance for each of these architectural paradigms to identify possible future designs. The leading five aspects of this novel work that compared multiple paradigms are as follows.

First, we model representative architectures based on state-of-the-art Simba, Tetris and UpMem architectures to represent CHA, NDP and PIM DNN accelerator paradigms based on a survey of each paradigm space, respectively.
Second, we use the latency, bandwidth and energy measurements of components from realized hardware architectures as inputs to the model, for a realistic comparison. 
This helps account for form factors and fabrication process limitations associated with each paradigm. 
Third, we use DNN layers from MLPerf \cite{mattson2019mlperf} to benchmark and identify the suitability of each architecture for different real-world applications under their respective latency and energy constraints.
Fourth, a detailed architectural cost model (based on Simba, Tetris, and UpMem) is used for each architecture, which considers the spatial and temporal data flow constraints between each component of the accelerator and its energy consumption. Further, the optimum mapping of the DNN to the detailed model, for least latency and energy consumption, is used for comparison.
Fifth, we perform sensitivity analysis on the detailed model to find architectural changes that benefit each application.

The main contributions of this work are:

\begin{itemize}[leftmargin=*]
\item We provide the first realistic and detailed comparison of state-of-the-art conventional, NDP and PIM paradigm-based DNN accelerators for their suitability for processing DNN applications. The models used in this work are based on hardware-realized designs.
\item Data-Reuse (\emph{DR}) of a DNN layer represents the number of times a word of input data participates in multiply-accumulate (MAC) operations. We show that for DNN applications with low data reuse, such as language and recommendation tasks, NDP architecture is $\sim$10$\times$ and $\sim$39$\times$ faster for inferencing than CHA and PIM architectures, respectively. 
\item CHA has $\sim$10$\times$ and $\sim$2000$\times$ lower latency than NDP and PIM architectures, respectively, for inferencing DNN applications with high \emph{DR} ($\sim$100~MACs/word) such as object detection and image classification.
Because CHA architecture can supply required data to a vast number of fast MAC units from its internal buffer without being affected by the narrow bandwidth between the accelerator and memory for such applications.
\item For applications with constraints on energy, such as mobile applications, PIM paradigm is better as it consumes the least energy ($\sim$2.7$\times$ and $\sim$21$\times$ less than CHA for Convolutional Neural Networks (CNN) and Fully Connected Layer (FCL) applications respectively) for DNNs. This is because energy for data transfer from memory to MAC units is significantly less than other considered architectures.
\item 
We identify architectural changes that can increase throughput, decrease latency and lower energy for ML applications with sensitivity analysis. Batching is universally beneficial for CHA in terms of throughput and energy, at the cost of slightly increased latency. However, we identify that batching can adversely affect BERT and DLRM in NDP and PIM systems. An increase in last-level-memory bandwidth can decrease the latency of CHA by 20$\%$ for CNN applications and linearly increase performance for FCL applications.
\item
Further, by sensitivity analysis, the following observations are made. (i) NDP can be redesigned to be 3$\times$ faster than CHA for CNN applications, thus making it the fastest paradigm for both CNN and FCL workloads. (ii) PIM is not suitable for CNN workloads when latency is the primary concern. (iii) PIM, with design changes, is more suitable for FCL workloads than CHA.
\end{itemize}

\section{Background: Software for DNNs}\label{backgroundSoftware}

Deep Neural Networks (DNNs) can be broadly classified into artificial neural networks and spiking neural networks. These DNNs usually have two phases, namely training and inference. In this work we analyze the design of accelerators used during the inference phase of the artificial neural networks. We have focused on inference as training, in general, is done infrequently. Inference, on the other hand, is done every time the application needs to perform a task. Also, latency, throughput energy, or power constraints are usually tighter for inference application scenarios than training, which require specialized hardware accelerators~\cite{surveyofhardware}.
The background related to the design of DNNs and the accelerators for these DNNs are discussed in the next sections.
\subsection{Inference} 
In DNNs, inference is performed by passing a high-dimensional input through a parameterized function. This parameterized function can be represented as a stack of neural network layers as shown in Eq.~\ref{DNNpipeline}. 
\begin{equation}\label{DNNpipeline}
  f(x) = f_{l-1}\circ f_{l-2}\circ ...\circ f_{2} \circ f_{1}(x)
\end{equation}
Here $x$ represents the high dimensional input.
Each of the functions ($f_{1},f_{2}, ..., f_{l-1}$) form a layer, and the output of each layer is passed only to the next layer in a uni-directional manner forming a network. Processing these layers on a traditional computer with general purpose CPU based on von Neumann architecture involves huge volumes of data transfer between the processor and the off-chip memory due to large network size and high number of memory accesses~\cite{surveyofhardware}. This limits the system's performance and increases the energy consumption during inference, making traditional systems highly inefficient for performing inference using DNNs. Furthermore, traditional systems suffer from two physical limitations as follows. (i) Due to the existing memory wall~\cite{memorywall}, where the processor's latency being much lower and internal bandwidth being higher than memory's latency, memory is unable to transfer sufficient data in time to the processing units. Hence the processing unit's utilization is significantly reduced, leading to lower performance. (ii) Traditional systems have complicated sub-units (modules of out-of-order core) which lead to significant power consumption, making them inefficient at processing DNNs. High power dissipation due to the power wall~\cite{powerwall}, i.e., comparable leakage and dynamic power in lower technology nodes, further put a limit on the performance of these processing units.  
These limitations have led to the shift away from traditional systems to the development of accelerators tailored toward DNNs~\cite{surveyofhardware}. These accelerators have orders-of-magnitude higher performance and power efficiency as compared to traditional computer systems~\cite{Simba2}. They mitigate the issues related to bottlenecks of traditional systems and also successfully exploit the deterministic flow of data in DNNs~\cite{chen2014diannao}.

\begin{figure}%
    \centering
    \subfloat[\centering Fully Connected Layer\label{Fig:FCL layer}]{{\includegraphics[width=0.5\linewidth]{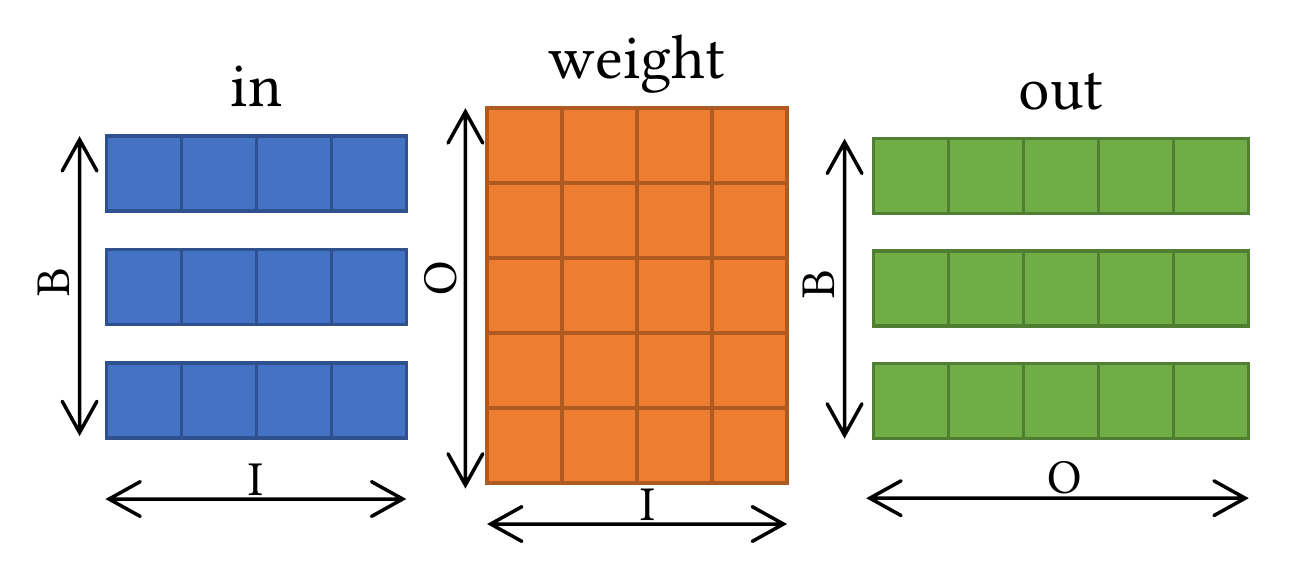} }}%
    \\
    \subfloat[\centering 2D Convolutional Layer - Sliding Window\label{Fig:CLlayer} ]{{\includegraphics[width=\linewidth]{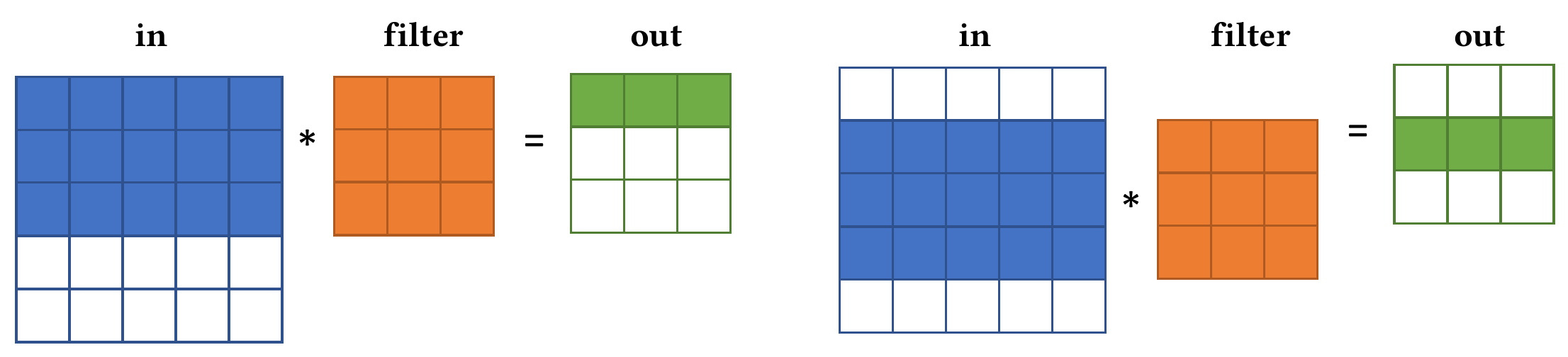} }}%
    
    \caption{Shape and operation involved in a layer of DNN}%
    \label{fig:layershape}%
    
\end{figure}

DNNs can have different types of layers such as Fully Connected Layer (FCL), Convolutional Layer (CL),  ReLU, sigmoid, max pooling, and batch normalization layers. In state-of-the-art DNNs, FCL and CL account for over 90\% of the operations~\cite{surveyofhardware}. Hence, most DNN accelerator designs are targeted towards accelerating these layers \cite{surveyofhardware}. In this work, we have also focused on FCL and the CL of the DNNs. 
\begin{equation}\label{FCLLayer}
  out_{b,o_{c}} = \sum_{i_{c}=0}^{I_{c}-1} in_{b,i_{c}} \times weight_{i_{c},o_{c}}\\
\end{equation}
A FCL performs matrix multiplication and can be represented as Eq.~\ref{FCLLayer} and Fig.~\ref{Fig:FCL layer} where $I$, $O$, and $B$ denote the number of input channels, number of output channels, and batch size (number of separate inputs that are being processed together), respectively. Further, in the equation, $i$, $o$, and $b$ denote the index of input channels, output channels, and batch, respectively. Similarly, a CL can be represented by the Eq.~\ref{CNNLayer} and Fig.~\ref{Fig:CLlayer}, where $F_h$ and $F_w$ are the height and width of the filter, $i$ and $j$ represent the row and column index of an element in the filter, $x$ and $y$ represent the index of an element in the feature map. Contrary to FCL, in CL, the filter is slid across the input and element-wise multiplied and accumulated to generate the output, as shown in Fig.~\ref{Fig:CLlayer}.
For CL and FCL, the inputs are reused for all output channels, and weights are reused for all input batches. This data reuse is also an important property that is exploited in the DNN accelerator designs to make them efficient. 


\begin{equation}\label{CNNLayer}
  \\
out_{b,o_{c},x,y} = \sum_{i_{c}=0}^{I_{c}-1} \sum_{i=0}^{F_{h}-1}\sum_{i=0}^{F_{w}-1}in_{b,i_{c},x+i,y+j} \times filter_{o_{c},i_{c},i,j_{c}}
\end{equation}

\subsection{Inference Benchmark}
There are a number of DNNs that have been proposed in the past decade. To effectively capture the state-of-the-art DNNs, the benchmark networks adopted in MLPerf Inference benchmark have been used in this work. MLPerf Inference benchmark is a benchmark suite made up of workloads that are reflective of real-world applications~\cite{mattson2019mlperf}.
Since our goal in this work is to perform a rigorous comparison among the state-of-the-art accelerators, we have used unique layers from MobileNet~\cite{howard2017mobilenets}, ResNet~\cite{resnet}, BERT~\cite{devlin2018bert}, and DLRM~\cite{dlrm} of MLPerf Inference benchmark for our analysis. 

\begin{table}[]
\centering
\caption{Index of unique layers of MobileNet, ResNet, BERT and DLRM}
\label{tab:indextable}
 \centering
  \includegraphics[width=0.8\linewidth]{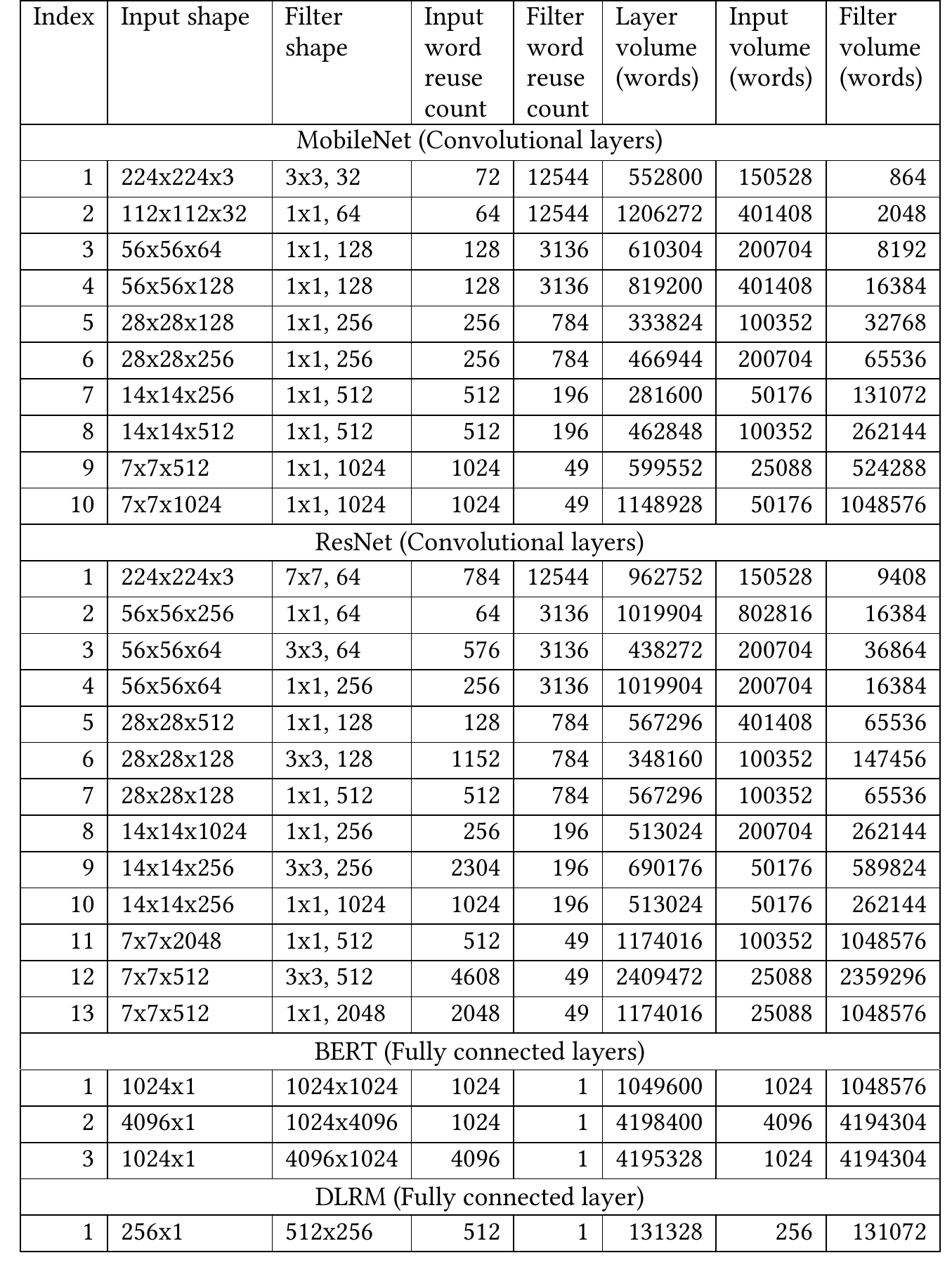}

\end{table}
MobileNet (MN) is designed for mobile and embedded vision applications and is designed for energy efficiency. ResNet (RN) is designed to be efficient when training with a deeper number of layers by alleviating the issues related to vanishing gradients using skip connections~\cite{resnet}. ResNet has improved object detection accuracy due to its extreme deep representations. BERT is a language representation model with applications in various tasks such as question answering and language inference. DLRM is used for recommendation tasks.
The 27 unique layers have been taken from these networks as shown in Table \ref{tab:indextable}. The index numbers and corresponding layers belonging to each NN used in this work are also shown. The table also shows the reuse potential of input and filter data words. DNN accelerators heavily exploit this data reuse by caching operands to reduce bottlenecks due to memory wall and improve energy efficiency \cite{surveyofhardware}. Further, \emph{layer volume} provides the minimum number of words of the NN layer that have to be moved between main memory and the compute units when the NN is processed in a layer by layer manner. In a pipelined setup with multiple tiled regions for compute, \emph{layer volume} is the number of words that needs to transferred to the next region \cite{Simba2}. Minimum data transfer is achieved when using data-stationary approaches where either the entire input, filter, or output is kept on-chip during the computation being done by a layer. \emph{Input volume} and  \emph{Filter volume} of a layer denote the minimum required size of the on-chip buffers to implement input and filter stationary approaches of computing, respectively.
In the following section, the design of different accelerator types and details related to data reuse have been explained. 

\section{Background: Hardware for DNNs}\label{backgroundhardware}
Traditional von Neumann computers have limits on system performance and energy efficiency when processing DNN applications. Also, most DNN applications have strict latency requirements, and traditional von Neumann computers cannot meet these requirements~\cite{surveyofhardware}. Further, traditional von Neumann computers are made up of separate processor and off-chip memory components~\cite{chen2016eyeriss}. Because of this separation, it takes $\sim$200$\times$ more energy to transport operands of an instruction from memory to the processor than the actual computation \cite{chen2016eyeriss,UpMem}. These limits have led to the development of computing platforms dedicated to processing DNN applications. 

Contemporary DNNs have 10 to 200 million compute operations per layer, as seen in Table \ref{tab:indextable}. A large fraction of these compute operations can be performed in parallel due to minimal data dependency between the output data words. This idea is heavily exploited in the design of DNN accelerators.
Over the years, many accelerators designs have been proposed for DNNs. Early designs \cite{copro1,copro2} were on-chip co-processors with limited compute capabilities. Conventional standalone DNN accelerators~\cite{standalone1, Simba2} are being designed with an understanding of the workloads where data flow is analyzed and utilized to reduce off-chip accesses and increase system performance and efficiency. We classify these as Conventional Hardware Accelerators (CHA). CHAs have massively parallel compute units that allow hundreds of Multiply-Accumulation (MAC) operations per cycle~\cite{Simba2}. However, these CHAs still suffer from the limitation of fetching data from off-chip memory, which is an energy-intensive operation.

To overcome these limitations, accelerators have been proposed which perform computations either near or in-memory, i.e., the computations are closely coupled to memory, alleviating the issue of off-chip memory accesses. We classify these accelerators as non-conventional hardware accelerators (NHAs)~\cite{tetris,UpMem}. NHAs are further subdivided into two categories.   
 
The first type, such as~\cite{UpMem} employs digital compute circuitry near the row buffers of 2D memory to minimize data movement. However, this architecture is limited to simple logic designs owing to the limited number of metal layers available in the process \cite{pimcasestudy}. The second type, such as~\cite{tetris}, exploits vertical integration of DRAM, and logic dies of 3D memories by placing the compute elements in logic dies. Accelerators, such as \cite{reram1,reram2}, utilize memory cells to perform computations using analog properties. In this work, we limit ourselves to the first two categories, namely CHAs and NHAs based on 2D and 3D memory.

\subsection{Conventional Accelerators}

Conventional hardware accelerators are either standalone systems or part of a general-purpose host system. In either case, the Last-Level-Memory (LLM) used is the traditional DDR DRAM, and the chips housing the MAC units are designed specifically for a set of DNN applications. Hence, these systems are primarily bottlenecked due to the memory wall \cite{surveyofhardware}. For these systems, it is essential to exploit the data reuse potential of NN applications to reduce off-chip data transfers. As shown in Fig.~\ref{fig:SimbaDetailed}, these systems use large on-chip buffers to hold the inputs, intermediate outputs, and filters while fast processing elements (PEs) perform computations using these data. A Network-on-chip (NoC) helps data transfer among PEs. Techniques such as multi-casting are used to deliver required data to all the PEs in parallel if needed~\cite{chen2016eyeriss,Simba2}. PEs have local buffers to reduce traffic on NoCs and also have specialized Multiply-Accumulate (MAC) structures to perform operations in DNN. The spatial arrangement of PEs allows for the splitting of the these operations by input and output channels across different PEs. While there are a number of CHAs \cite{alwani2016fused, chen2014diannao, chen2016eyeriss, han2016eie, parashar2017scnn, qadeer2013convolution, sharma2018bit, sijstermans2018nvidia}, in this work, we have used the state-of-the-art, hardware realized Simba accelerator as a representative architecture for CHAs.

\subsubsection{SIMBA}

\begin{figure}[h]
  \centering
  \includegraphics[width=\linewidth]{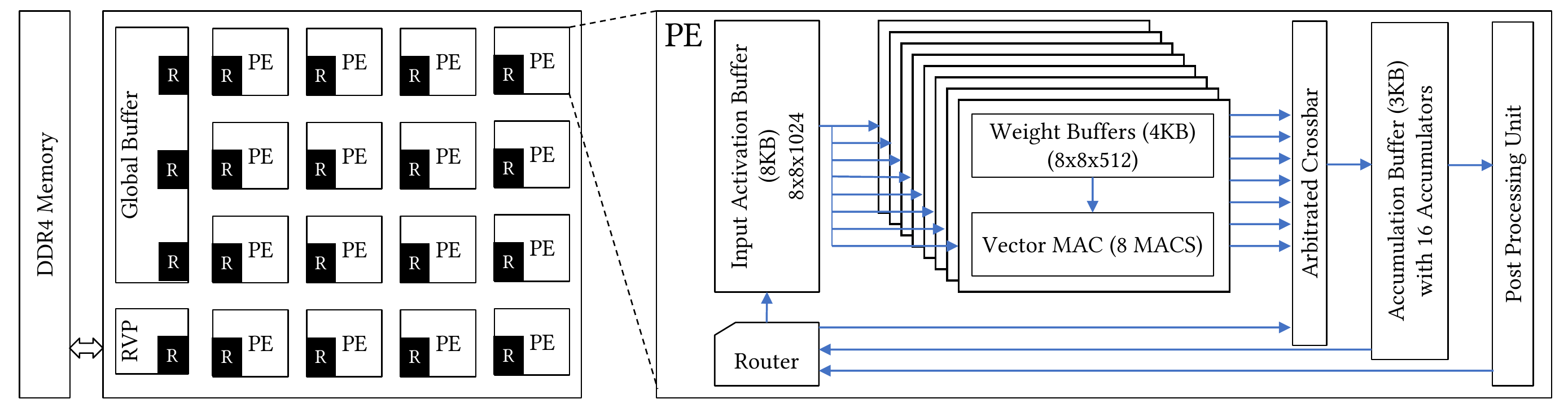}

  \caption{Simba and its PE architecture}

  \label{fig:SimbaDetailed}
\end{figure}

Simba \cite{Simba2} is a highly scalable chipset-based architecture with the baseline system consisting of one chiplet. The chiplet consists of resources for inter-chiplet communication, a Global Buffer, an array of 16 PEs connected together using an NoC and a RISC-V processor (RVP) to co-ordinate the compute tasks, as shown in Fig.~\ref{fig:SimbaDetailed}.
Each PE has a router interface, and buffers for weights (filters), input, and output (\emph{Accumulation Buffer}). Further, each PE has eight vector MACs, and each vector MAC is capable of Multiplying and accumulating eight input words with eight weight words to form a single output word. The \emph{Accumulation Buffer}, with the help of separate accumulators, accumulates current results with previous partial results stored in the \emph{Accumulation Buffer}. Results transferred from other PEs over the NoC can also be accumulated. The generated output is then post-processed (ReLU, Pooling, Scaling) or directly transferred out of the PE via the NoC.

\subsection{Non Conventional Hardware Accelerators}
 
Conventional applications have dynamic paths for execution and require techniques such as branch predictors and prefetchers to improve performance. However, most DNN applications have a pre-determined fixed pattern for computing. The DNN applications can be decomposed into simple parallel operations that can be performed using simple compute units in a deterministic manner~\cite{parashar2019timeloop}. Further, the throughput can be increased by performing these simple compute operations in parallel. However, a DNN accelerator system having separate off-chip memory and compute units still introduces scalability challenges due low bandwidth and high data access energy of off-chip memory. Hence NHAs were introduced to overcome this limitation. In most NHAs, the memory and the compute units are tightly integrated, alleviating the issue of memory wall due to the high internal bandwidth. NHAs can further be sub-categorized into two categories:

\begin{figure}[htpb]
    \centering
    \subfloat[\centering NN Engine within a vault of Tetris\label{fig:NHATetris}]{{\includegraphics[width=0.55\linewidth]{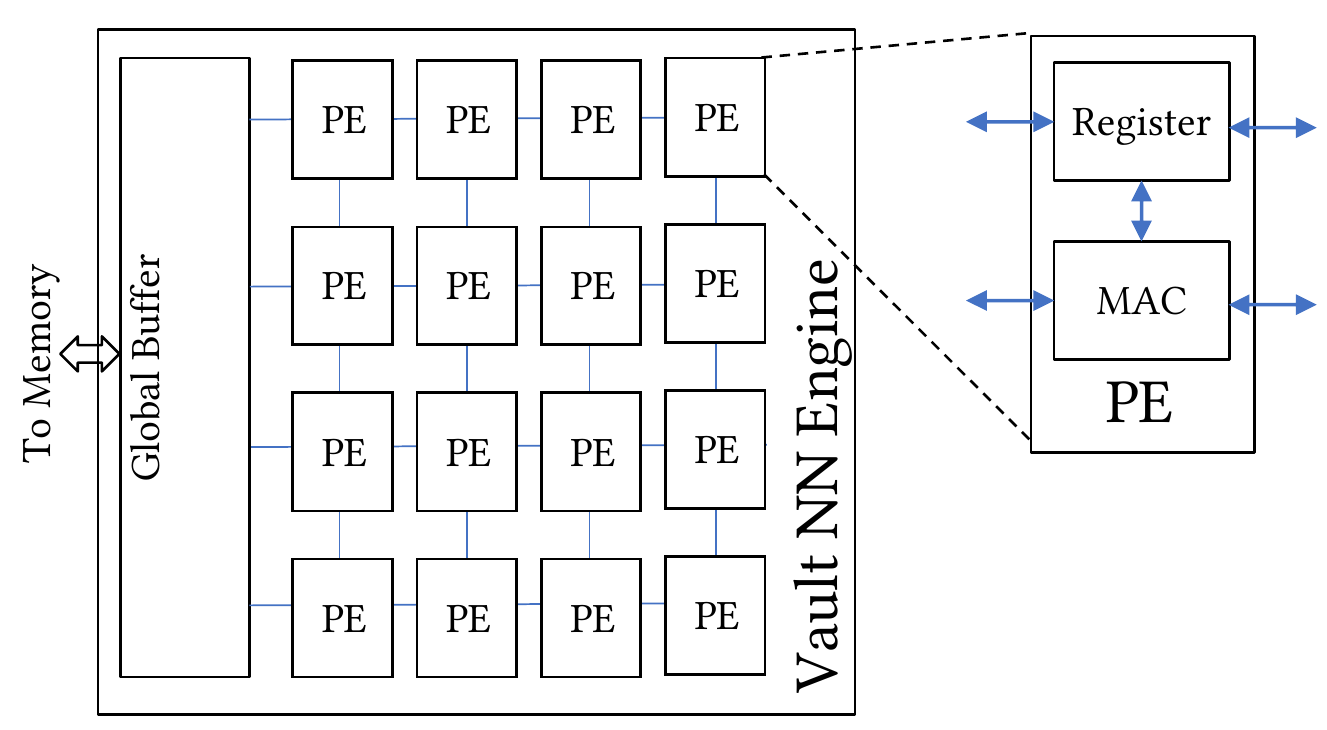} }}%
    \\
    \subfloat[\centering A Single PIM Chip in UpMem \label{fig:NHAUpMem}]{{\includegraphics[width=0.55\linewidth]{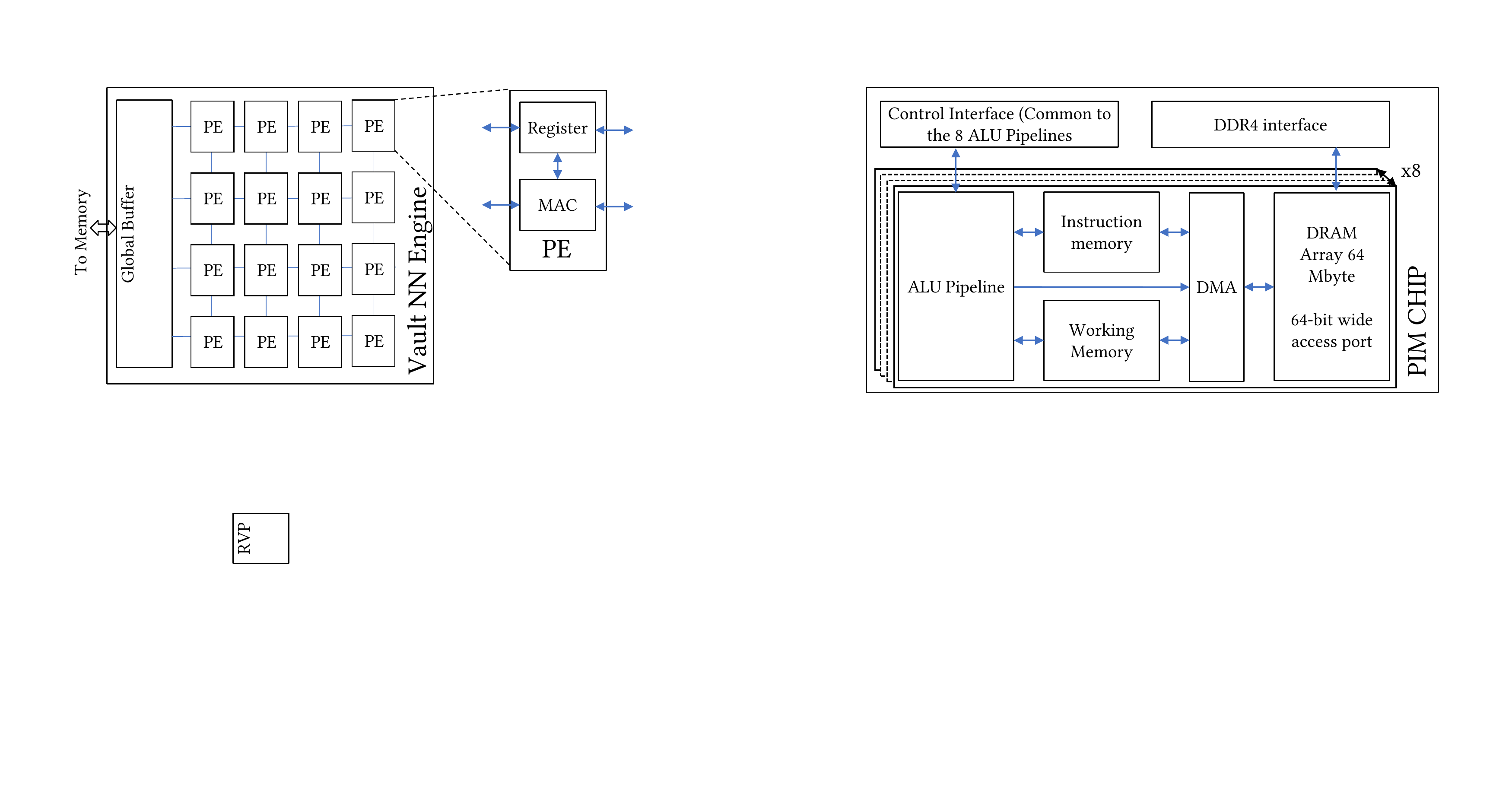} }}%
   
    \caption{Architecture of basic building block of 3D and 2D NHA}%

    \label{fig:NHA}%
\end{figure}
\subsubsection{3D Memory - NDP}

Fig.~\ref{fig:Architectures}c shows Tetris \cite{tetris}, a state-of-the-art 3D memory-based DNN accelerator. In Tetris, the memory dies of the Hybrid-Memory-Cube (HMC) are stacked on top of the logic layer. The dies are spatially split into sixteen parts, as shown in the figure, and the parts that are stacked on top of each other are connected via through-silicon vias (TSVs) to form a vault. Each vault operates independently and contains a separate memory controller on the logic die. 
In an HMC, each vault has $3.5$ $mm^2$ of unused area in the logic layer, this area is utilized for dedicated compute units and associated resources such as buffers for performing DNN acceleration in Tetris. As shown in Fig.~\ref{fig:NHATetris}, each vault has a single instance of an accelerator with its own global buffer and PE array. The data stored in the same \emph{vault} is fetched with the help of the memory controller, while data stored in another vault has to be fetched over the inter-vault router. Since the computation is performed separately to memory die and the compute sites are near where data resides, this paradigm is known as Near-Data-Processing (NDP).

This category of NHA uses 3D memory, and for this work, we have used Tetris~\cite{tetris}, which is the state-of-art 3D memory-based accelerator for DNNs, for our analysis.
\subsubsection{2D Memory - PIM}

Fig.~\ref{fig:Architectures}e shows an abstract hardware model of UpMem, a state-of-the-art 2D memory-based DNN accelerator. UpMem is designed to work alongside a host system where the UpMem DIMMs act as both regular memory and compute accelerators. The DIMMs are constructed of PIM chips, as shown in Fig.~\ref{fig:NHAUpMem}. Each PIM chip has memory arrays that are connected to the host via a standard DDR interface. Further, each PIM chip has a small local buffer and dedicated compute units with access to memory arrays over DMA.
 This arrangement provides large bandwidth to the compute units. However, inter-chip communication still relies on the standard DDR interface and is costly in terms of latency and energy.

To get insights into the different accelerators, we will compare Simba, Tetris, and UpMem. A detailed and quantified comparison of the architectural parameters for each of these is given in Section~\ref{archConstraints}.
In the next section, we discuss the evaluation setup used to compare the architectures.

\section{Representative models for DNN accelerator paradigms} \label{representativemodel}
A fine-grained analysis of all DNN accelerators belonging to CHA, NDP, and PIM is beyond the scope of this work. Hence we construct representative models for each paradigm with the best in class parameters from data-agnostic works. In data-agnostic processing of NN, the actual value of inputs do not change the inference time. Further, the parameters are measured only from hardware-based works. Furthermore, for the spatial organization of the components representing each paradigm, we use Simba, Tetris and UpMem for CHA, NDP and PIM. They are state-of-the-art hardware-based designs, and hence the architectural parameters inscribing the aspects are constrained by the real-world characteristics of each paradigm. Further, Simba, Tetris and UpMem are data-agnostic designs with the most best in class parameters as compared to other relevant works in each paradigm, as shown in Table~\ref{tab:repres}, hence are good representatives of the paradigm space.

\begin{table}[h]
\centering
\caption{Representative hardware accelerators and relevant related works from each paradigm}
\label{tab:repres}
\resizebox{\textwidth}{!}{%
\begin{tabular}{|l|l|l|l|}
\hline
Representative Accelerator & Simba \cite{Simba2} (CHA)                                                                                                                                   & Tetris \cite{tetris} (NDP)                                                                              & UpMem \cite{UpMem} (PIM)                                                                                 \\ \hline
Related works              & \cite{alwani2016fused, chen2014diannao, chen2016eyeriss, han2016eie, parashar2017scnn, qadeer2013convolution, sharma2018bit, sijstermans2018nvidia} & \cite{ azarkhish2017neurostream, yin2018parana, wang2018exploiting, das2020nzespa, park2021high} & \cite{ hajinazar2021simdram, kwon202125, damov, kim2020trim, lee20221ynm, he2020newton, roy2021pim} \\ \hline
\end{tabular}%
}
\end{table}

\subsection{Architectural parameters}

For an accurate evaluation of DNN architectures, Oliveira et al. \cite{damov} and G{\'o}mez-Luna et al. \cite{gomez2021benchmarking} have shown that data storage and data movement bottlenecks must be modeled along with the compute capabilities of the DNN accelerators. Parashar et al. \cite{parashar2019timeloop} and Wu et al. \cite{wu2019accelergy} have shown that mapping a DNN to the DNN accelerator has to be optimized, based on the accelerator's structure and based on latency and energy aspects of the individual components of the accelerator, to achieve the lowest latency and energy for execution. In this work, the mapping is optimized for least latency followed by least energy, and such a mapping is referred to as \emph{optimal mapping}. Hence the relevant parameters are as follows. (i) Count and latency of MAC units. (ii) Working memory size (iii) Data bandwidth: bandwidth between different PEs, bandwidth between working memory and PEs, and bandwidth between DRAM and working memory. (iv) MAC compute energy (v) Data access and data transfer energies from each data storage level. The specific values of parameters and their relationships are explained in Section \ref{archConstraints}.

\section{Tools used for Analysis and Experimental Setup }

\begin{table}[h]
\centering
\caption{Summary of Experiments}
\label{tab:summaryofExperiments}
\resizebox{0.8\linewidth}{!}{%
\begin{tabular}{|ll|}
\hline
\multicolumn{1}{|c|}{\textbf{Experiment}}                                                                                                                                                         & \multicolumn{1}{c|}{\textbf{Figures}}                                    \\ \hline
\multicolumn{2}{|c|}{\begin{tabular}[c]{@{}c@{}}Note: The metrics for comparison are obtained for each experiment where mapping \\is optimized for least latency followed by least energy for Simba-like,  Tetris-like \\and UpMem-like systems.\end{tabular}} \\ \hline
\multicolumn{1}{|l|}{\begin{tabular}[c]{@{}l@{}}Baseline – Inference of MobileNet, ResNet, BERT and DLRM\end{tabular}}                                                                         & \begin{tabular}[c]{@{}l@{}} \ref{fig:CNNlatency}, \ref{fig:CNNUtilization}, \ref{fig:CNNEnergy}, \ref{fig:CNNlayerEnergysplit},  \ref{fig:FCLTime},          \ref{fig:FCLUtilization},       \ref{fig:FCLEnergy},         \ref{fig:FCLEnergySplit} \end{tabular}                          \\ \hline
\multicolumn{1}{|l|}{\begin{tabular}[c]{@{}l@{}}Batching –  \\The workload inputs are batched at 1$\times$, 2$\times$, 4$\times$ and 8$\times$.\end{tabular}}                                                                 & \begin{tabular}[c]{@{}l@{}}\ref{fig:BatchingCNNTime},       \ref{fig:BatchingCNNUtilization},     \ref{fig:CNNBatchingEnergy}      , \ref{fig:BatchingCNNEnergyPerCompute},      \\ \ref{fig:BatchingFCLTime},            \ref{fig:BatchingFCLUtilization}, \ref{fig:BatchingFCLEnergy},          \ref{fig:BatchingFCLEnergyPerCompute}  \end{tabular}                          \\ \hline
\multicolumn{1}{|l|}{\begin{tabular}[c]{@{}l@{}}Last level memory bandwidth – The last level memory\\ bandwidth is increased by 1$\times$, 2$\times$ and 4$\times$.\end{tabular}}                                  & \begin{tabular}[c]{@{}l@{}}  \ref{fig:ExternalBWCNNTime},   \ref{fig:ExternalBWCNNUtilization},   \ref{fig:ExternalBWFCLTime},   \ref{fig:ExternalBWFCLUtilization} \end{tabular}                          \\ \hline
\multicolumn{1}{|l|}{\begin{tabular}[c]{@{}l@{}}Maximum usable MAC units- The total count of MAC \\ units for each system is increased to 100 thousand.\end{tabular}}                           & \begin{tabular}[c]{@{}l@{}}   \ref{fig:maxMACCNN}  \ref{fig:maxMACFCL},   \ref{fig:maxMACTime}\end{tabular}                          \\ \hline
\multicolumn{1}{|l|}{\begin{tabular}[c]{@{}l@{}}Buffer size and layout- The buffer size of each \\ system is increased to 1$\times$, 2$\times$ and 4$\times$.\end{tabular}}                                                       & \begin{tabular}[c]{@{}l@{}}\ref{fig:BufferSize},\ref{fig:BufferLayoutTime},\ref{fig:BufferLayoutPercompute}\end{tabular}                          \\ \hline
\end{tabular}}
\end{table}

The detailed models of the accelerators are implemented based on the designs described in  Section~\ref{backgroundhardware} and the parameters in Section~\ref{archConstraints}. Timeloop~\cite{parashar2019timeloop}, Accelergy~\cite{wu2019accelergy}, and Cacti~\cite{muralimanohar2009cacti} are used to perform detailed analysis. 
The per-layer optimal mapping of DNN to hardware is found using Timeloop.

Accelergy and Cacti are used to obtain the area and energy of PEs, SRAM buffers, and DRAM using the 45nm model. When calculating the total energy of DRAM, only the access energy of DRAM is considered for all three models.
Based on DNN's mapping and usage, timing, and energy associated with each accelerator component, the execution time and energy consumption are obtained. The summary of experiments carried out is shown in Table~\ref{tab:summaryofExperiments}.

\subsubsection*{Timeloop}
It allows the modeling of DNN accelerators and finds the optimal mapping of the workload to the accelerator based on user specified optimization criteria such as least latency or energy. The tool provides abstractions for defining compute units, memories across various levels, and communication links. Further, dataflow constraints, bandwidth limitations, and utilization constraints can be placed upon these three elements. The main usage of Timeloop in this work is to provide optimal mappings of DNN layers to architectural models. Timeloop's mapper constructs mapspace and evaluates the quality of each mapping with the cost model provided by Accelergy. The optimal mapping is obtained using an iterative search in the mapspace based on optimization criteria. The mappings are optimized for least latency followed by least energy in this work~\cite{parashar2019timeloop}.

\subsubsection*{Accelergy} 
It allows fast and accurate cost modeling, in terms of latency, area and energy, of each of the units present in DNN accelerator hardware. Similar to Timeloop, Accelergy provides abstractions to define compute units and various memory levels. Upon instantiating these abstractions, the corresponding area and energy values are obtained by Accelergy and are used by Timeloop for model mapping. In this work, Accelergy uses external plugins (Cacti and Aladdin\cite{shao2014aladdin}) to determine the area and energy of buffers and MACs.

\subsubsection*{Cacti and Aladdin
}
Cacti is a tool for modeling the dynamic power, access time, area, and leakage power of caches and other memories. 
Cacti is used to model the area and energy of buffers and memories in our work, with access time constraints. In this work, the 45 nm model is used and buffer size is based on values from Table \ref{tab:constraints}.

Aladdin allows for specifying scalable compute energy cost per bit.

In the next section, we discuss detailed architectural parameters that have been used for modeling of accelerators.

\section{Detailed Architectural Parameters}\label{archConstraints}
The values of architectural parameters used in this work are given in Table \ref{tab:constraints}. Dependent energy values are derived from CACTI based on values in Table \ref{tab:constraints}.


\begin{table}[htpb]

\centering
\caption{Summary of architectural constraints} 

\label{tab:constraints}
\resizebox{0.6\linewidth}{!}{%
\begin{tabular}{|l|c|c|c|}
\hline
\textbf{Metric/Constraint}                                                           & \textbf{CHA} & \textbf{NDP}                                             & \textbf{PIM} \\ \hline
MAC units                                                                 & 1024 \cite{Simba2}          & 256 \cite{tetris,damov}                                                         & 128 \cite{fimDRAM, UpMem}            \\ \hline
MAC latency                                                               & 1 ns \cite{Simba2}           & 2 ns \cite{tetris}                                                       & 40 ns  \cite{UpMem,gomez2021benchmarking}        \\ \hline
\begin{tabular}[c]{@{}l@{}}MAC compute \\ energy\end{tabular}             & 0.2 pJ/bit \cite{UpMem}     & 0.2 pJ/bit \cite{tetris}                                                 & 0.4 pJ/bit \cite{UpMem}    \\ \hline
Inter-PE bandwidth                                                            & 70 GB/s \cite{Simba2}        & 16 GB/s \cite{hbm2} & 12.8 GB/s \cite{gomez2021benchmarking}     \\ \hline
\begin{tabular}[c]{@{}l@{}}Working memory\\ size\end{tabular}             & 3 MB \cite{damov}           & 2 MB  \cite{tetris}                                                      & 512 KB \cite{gomez2021benchmarking}        \\ \hline
\begin{tabular}[c]{@{}l@{}}Working memory\\ bandwidth\end{tabular}        & 70 GB/s \cite{Simba2}       & 6.4 GB/s \cite{tetris}                                                   & 800 MB/s \cite{gomez2021benchmarking}      \\ \hline
\begin{tabular}[c]{@{}l@{}}DRAM-Compute\\ bandwidth\end{tabular}          & 25.6 GB/s \cite{jdecstd}      & \begin{tabular}[c]{@{}c@{}}16 GB/s\\ per vault \cite{tetris}\end{tabular} & 102 GB/s \cite{gomez2021benchmarking}      \\ \hline
\begin{tabular}[c]{@{}l@{}}DRAM-Compute\\ data access energy\end{tabular} & 46 pJ/bit  \cite{UpMem}    & 4.2 pJ/bit \cite{tetris}                                                 & 2.3 pJ/bit \cite{UpMem}     \\ \hline
\end{tabular}}
\end{table}

\subsubsection*{MAC Unit}
The MAC unit consists of circuits used to perform MAC operations and a register file with 16 entries. CHA architecture that we have considered in this study has 1024 MAC units, with each unit operating with a latency of 1 ns. Further, the compute energy is 0.2 pJ/bit. In NDP, MAC units drop to 256, and the latency doubles to 2 ns due to a stricter thermal constraint of 10W and area constraint of 3.5 $mm^2$ per vault. Further, the energy cost per bit of computing is 0.2 pJ/bit. CHA and NDP have similar energy costs for computing regardless of their difference in operating speed as state-of-the-art CHA have fused vector MACs that reduce writes. PIM system has only 128 MAC units across 16 PIM chips, with each MAC operating at a latency of 40 ns. The extreme slowdown is due to the DRAM process used for the MACs. Further, energy cost per bit also doubled to 0.4 pJ/bit due to the same reason. In CHA, the MAC count is limited by the available memory bandwidth, while in NDP and PIM, thermal, area, and process constraints limit the MAC count.

\subsubsection*{Working Memory}
All three paradigms have working memory to avoid redundant data fetches to the LLM. In CHA, the working memory is formed by the global buffer and the local buffers at each PE. However, due to relatively smaller configuration, NDP and PIM have only one buffer acting as the working memory between MAC units and the last-level-memory. Due to its memory bandwidth constraint and relaxed area constraints, CHA systems have large buffers (3 MB combined) to exploit the data-reuse potential of the workload. On the other hand, NDP and PIM systems have smaller buffers (2 MB and 512 KB, respectively) due to high bandwidth and relatively low access energy at the last-level memory.

\section{Metrics for Evaluation}

Architectures for ML inference operate under different operational \emph{scenarios} and can be compared using various metrics. Real-time ML applications operate in a \emph{single-stream scenario} where each input needs to be processed, and decisions based on it need to be taken before processing the following input.
In \emph{multi-stream scenario}, the objective is to handle as many parallel streams as possible without breaking the latency constraints required by the application. Surveillance cameras and sensor data processing in autonomous vehicles are excellent examples of this scenario. In the \emph{offline scenario}, the throughput of processing inputs is more important than the latency constraint and assumes that all the inputs are available from the start. Hence in this scenario, power is more important than energy. \emph{Server scenario} assumes the server setting where inputs arrive for inference at various rates. Therefore, the requests for processing the inputs will come at different times, creating a hybrid scenario between multi-stream and offline. Here the requests follow a Poisson distribution.

MLPerf utilizes mainly three metrics for performance comparison. (i) Latency: It measures the time it takes for an input to be fully inferenced. In our work, we measure the latency of each unique layer in different ML workloads in wall clock time (ns). Layer-wise measurement helps increase the coverage of ML workloads in MLPerf, as they are constructed of these unique layers in various combinations. The latency of each workload can be calculated by adding up the individual layer latencies. (ii) Throughput:  it is measured as layers per second in this work as batching is performed at layer level \cite{Simba2}. Measuring throughput is especially important for offline settings when multiple inputs are available for processing in parallel. (iii) Energy: MLPerf uses system energy per stream as a metric for comparing energy efficiency of systems. Latency and energy can be used to derive power values.

Apart from these metrics, ML accelerator designers are interested in the utilization ratio of all available MACs. This ratio informs the designer to understand compute resource requirements when processing each layer and how the architecture can be modified to suit the application better.

\section{Evaluation Results}
In this section we discuss the detailed analysis and comparisons across the three different architectures. The layer shapes respective to the index used in the Figures are given in Table~\ref{tab:indextable}.
\subsection{Convolutional Neural Network (CNN)}
\begin{figure}[htpb]
    \centering
    \subfloat[\centering Latency per layer\label{fig:CNNlatency}]{{\includegraphics[width=0.75\linewidth]{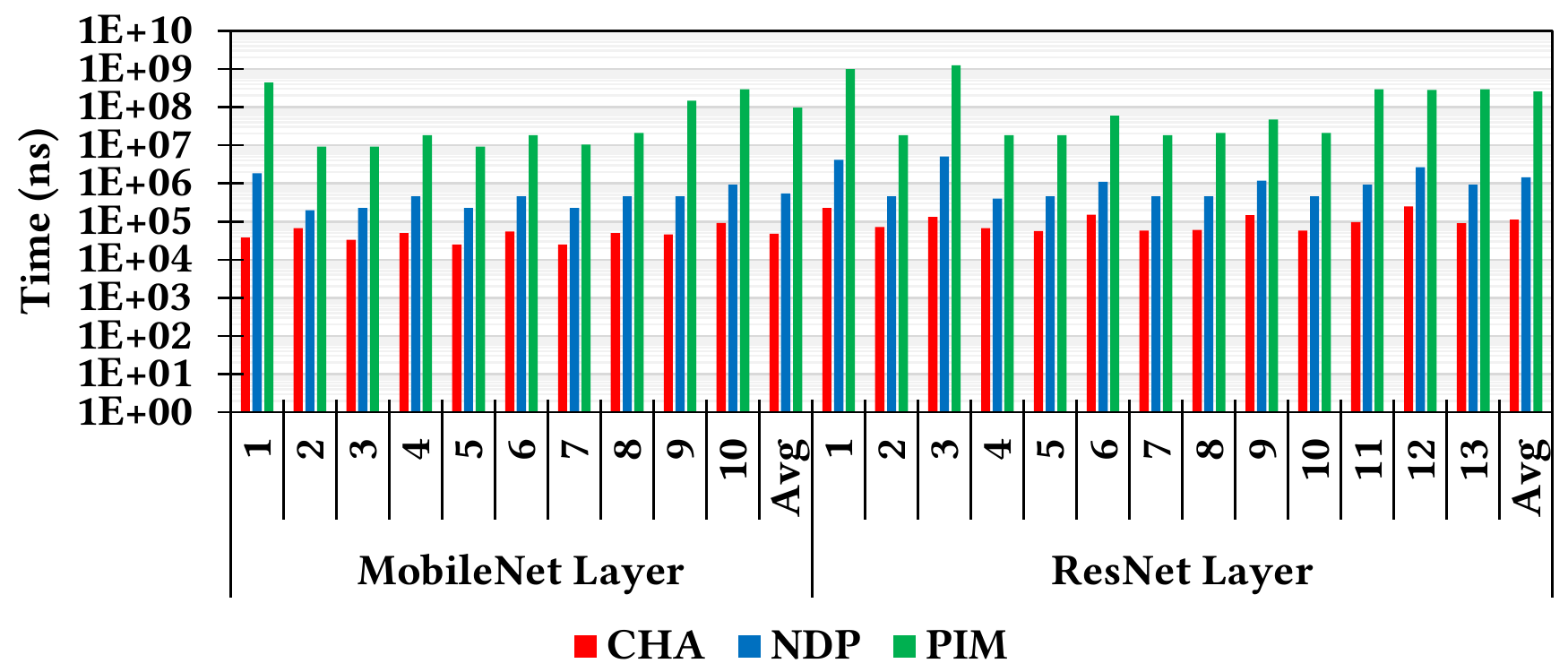} }}%
    \\ 
    \subfloat[\centering Layer level utilization of MAC\label{fig:CNNUtilization}]{{\includegraphics[width=0.75\linewidth]{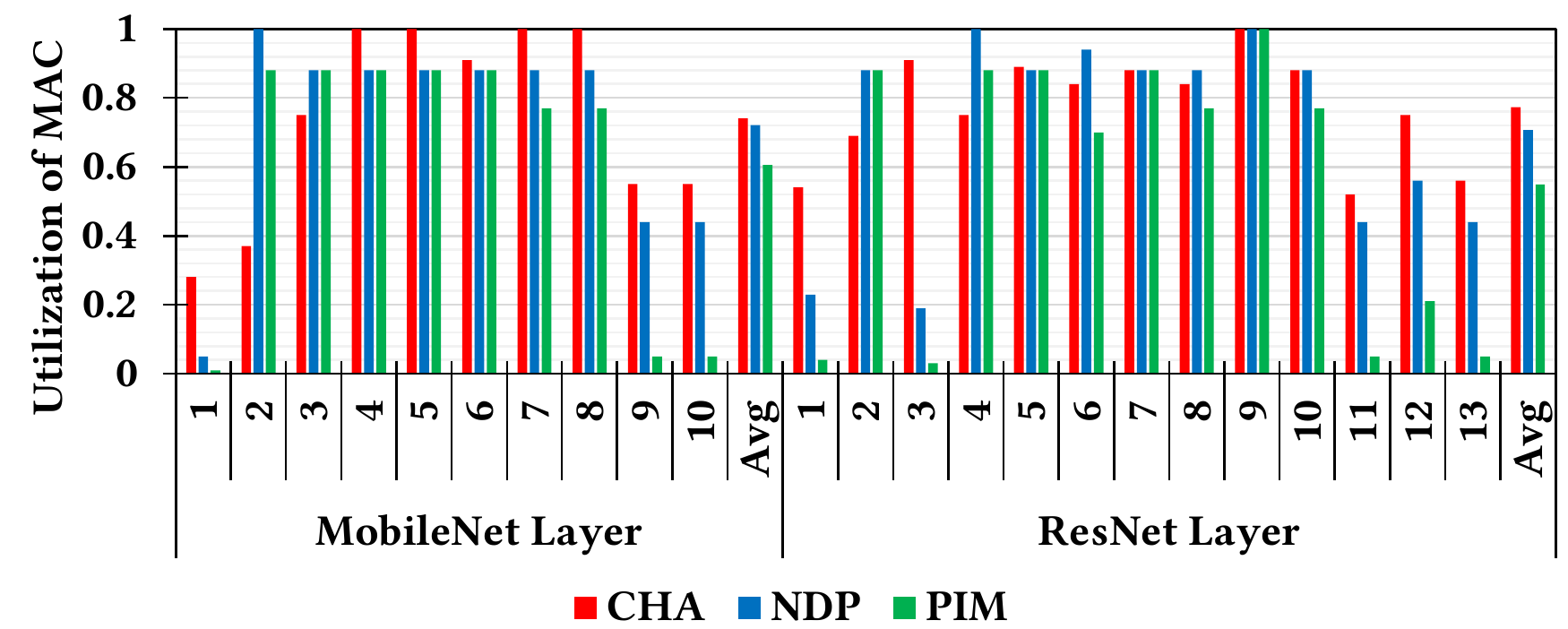} }}%
    \caption{Latency and utilization of MAC of CHA, NDP and PIM for CNN workloads for MobileNet and ResNet}%
    \label{fig:CNNlatency2}%
\end{figure}

In Fig.~\ref{fig:CNNlatency}, the average time taken by various architectures to process each layer of MobileNet (MN) and ResNet (RN) is shown. On average, CHA is $11\times$ and $2029\times$ faster as compared to NDP and PIM for MobileNet. Also,
CHA is $13\times$ and $2272\times$ faster as compared to NDP and PIM for ResNet.
For latency-sensitive ML vision applications that use CNN, CHA is preferable over NDP and PIM. This makes CHA ideal for use in the \emph{single stream} scenario where queries are generated sequentially and the 90$^{th}$ percentile latency is the biggest concern. It also makes it suitable for use in \emph{multi stream scenario} as it can incorporate more streams for a given latency bound. This performance gap further increases for the initial and the final layers. CHA becomes $21\times$ and $5600\times$ faster as compared to NDP and PIM for these layers. This variance is due to the skewed shape of these initial and final layers, making it less optimal to map the layers to the distributed architecture of NHAs.
The performance gap can be attributed to three main factors. (i)~Latency and the number of parallel MAC operations - CHA and NDP can complete 320 and 40 MAC operations in the time PIM performs 1 MAC operation. (ii) CHA has only 1/10$^{th}$ bandwidth of the NHAs, but data reuse in CHA is better exploited using the large on-chip buffer to compensate for the limited bandwidth, as shown in Section \ref{sensitibuffersize}. (iii) Due to the monolithic spatial architecture of CHA, it allows for higher utilization even with a batch size of one. As compared to NDP and PIM, CHA is able to better utilize its MAC units by 4\% and 17\% respectively, as shown in Fig.~\ref{fig:CNNUtilization}.

\begin{figure}[htpb]
  \centering
  \includegraphics[width=\linewidth]{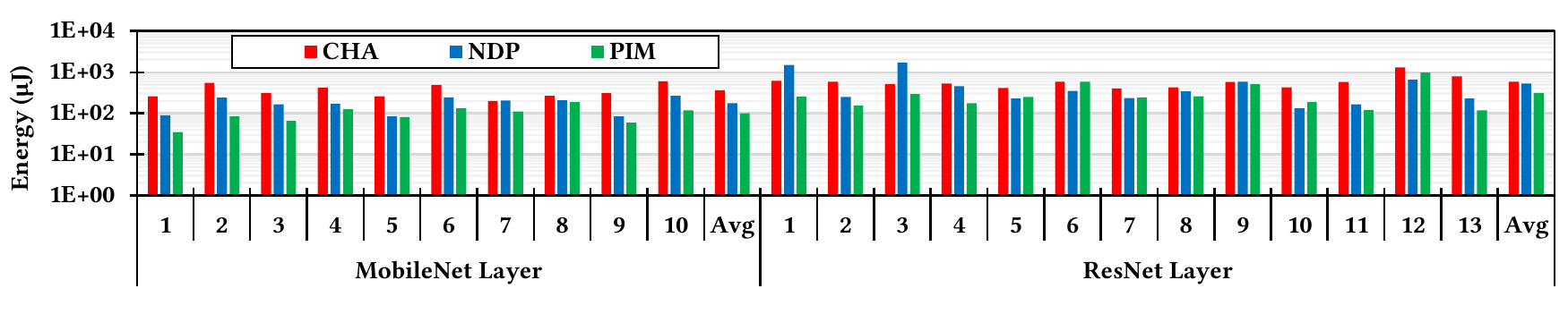}

  \caption{Energy consumed by each layer of MobileNet and ResNet}
 
  \label{fig:CNNEnergy}
\end{figure}

\begin{figure}[htpb]
  \centering
  \includegraphics[width=\linewidth]{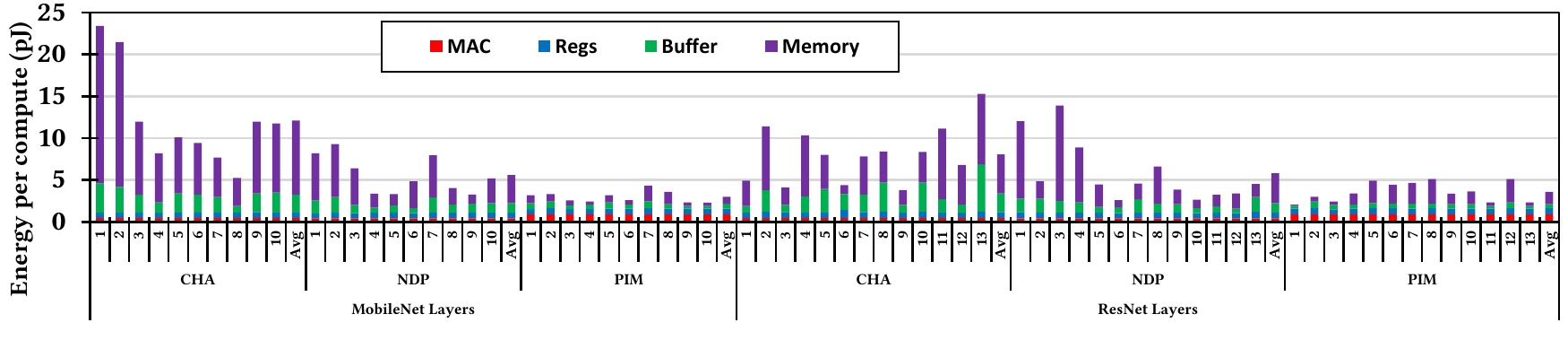}

  \caption{Breakup of energy per layer for each compute of MobileNet and ResNet}
 
  \label{fig:CNNlayerEnergysplit}
\end{figure}

Fig.~\ref{fig:CNNEnergy} shows the total energy consumed in each layer of MobileNet and ResNet, respectively. Fig.~\ref{fig:CNNlayerEnergysplit} show the split of energy consumed by individual parts of the system per computation.
PIM is the most energy-efficient among all designs. This can be largely attributed to no off-chip communication required for PIM. PIM consumes 1.7$\times$ and 3.6$\times$ lower energy than NDP and CHA for MobileNet. In CHA and NDP architectures, the data access to the LLM accounts for 73\% and 60\% of the total compute energy, while LLM access energy is only 27\% of the total energy per compute in PIM. CHA consumes the highest energy due to off-chip memory accesses. In NDP, the last level memory access energy is 2$\times$ more than that of PIM. This can be attributed to the TSVs used in NDP which are not present in PIM.

For ResNet, the difference in energy per compute for CHA and NDP narrows to 1.6$\times$ and 1.9$\times$, respectively, compared to PIM. For CHA, due to the higher reuse potential of ResNet compared to MobileNet, as shown in Table \ref{tab:indextable}, the number of LLM access is halved. Irrespective of this, the buffer energy consumption is quite large. NDP is not able to translate the higher data reuse potential of ResNet due to smaller buffers and distributed architecture, leading to higher energy consumption. 

In the next section, we discuss the analysis done for fully connected networks.

\subsection{Fully Connected Layers (FCL)}

\begin{figure}[htpb]
    \centering
    \subfloat[\centering Latency per layer for BERT and DLRM \label{fig:FCLTime}]{{\includegraphics[width=0.8\linewidth]{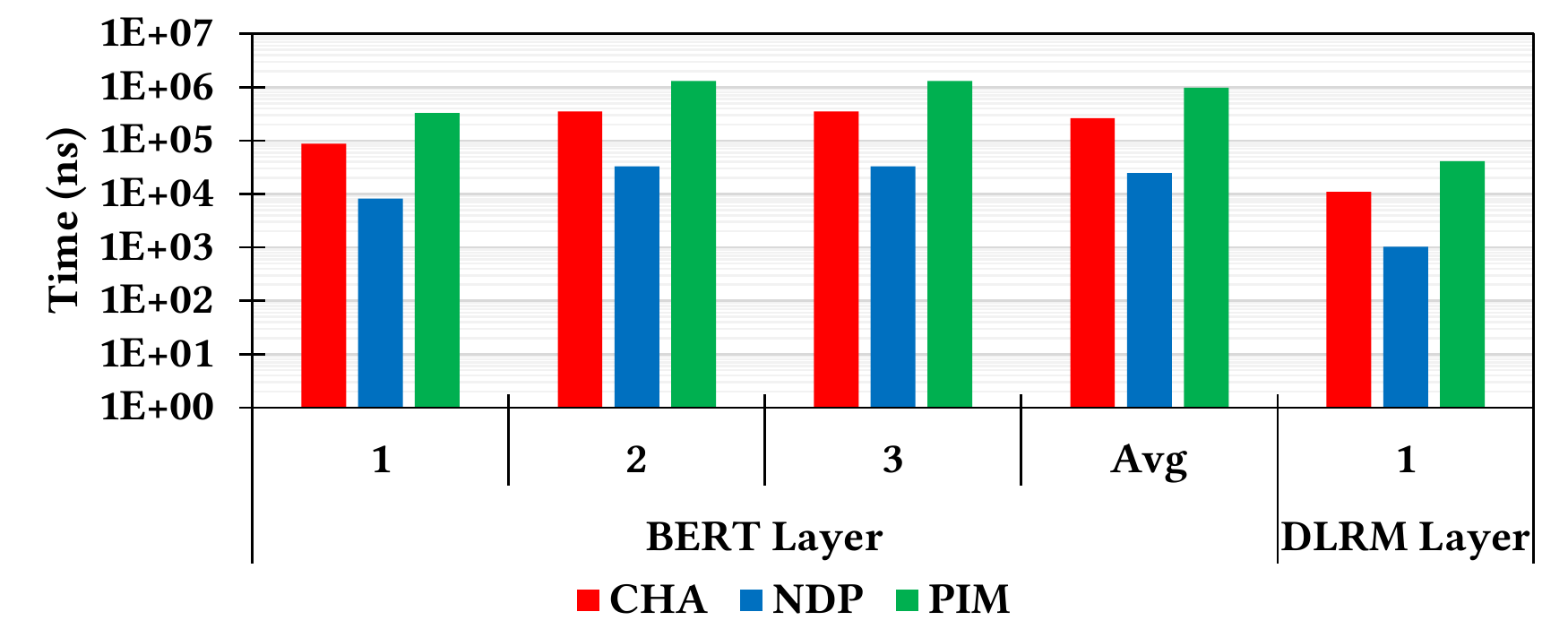} }}%
    \\
    \subfloat[\centering Utilization of MACs per layer for BERT and DLRM \label{fig:FCLUtilization}]{{\includegraphics[width=0.8\linewidth]{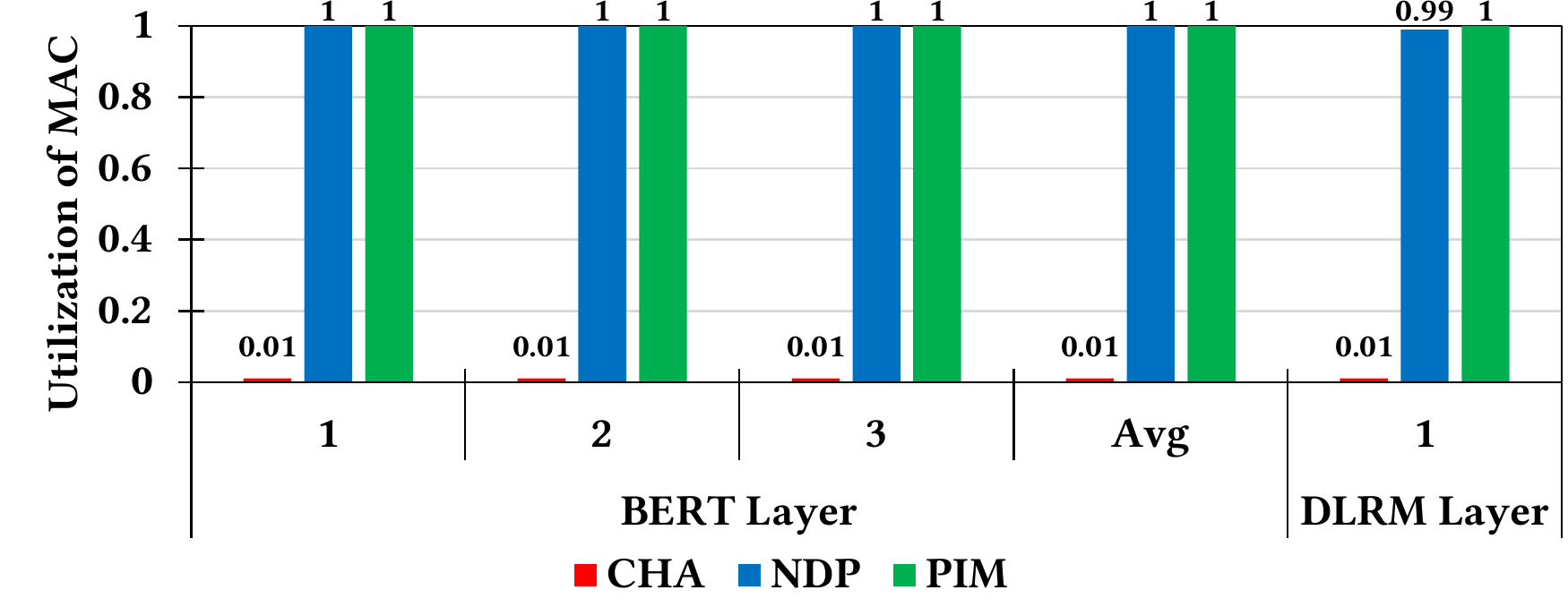} }}%
    \caption{Latency and utilization of MAC of CHA, NDP and PIM for FCL workloads}%
    \label{fig:FCLTime2}%
\end{figure}

Fig.~\ref{fig:FCLTime} shows the time taken by each FCL of BERT and DLRM, and  Fig.~\ref{fig:FCLUtilization} shows the MAC utilization of CHA, NDP and PIM for BERT and DLRM. For FCL layers, NDP is 11$\times$ and 40$\times$ faster than CHA and PIM. Unlike CNN, the data reuse potential is negligible, i.e., 1 compute/data word for FCL - leading to a severe LLM bandwidth bottleneck in CHA. This leads to only 1\% MAC utilization as shown in Fig~\ref{fig:FCLUtilization}. Contrary to CHA, in NDP, the utilization is 100\% as each vault has sufficient bandwidth to supply the required data to compute logic. Even though each MAC in NDP is working at half the frequency of CHA - the higher number of MACs working together compensates for the lower frequency leading to better performance compared to CHA. PIM has 100\% utilization same as NDP, but PIM is 20$\times$ slower because the logic in PIM is built using DRAM process, which is inherently slow, as explained in Section~\ref{backgroundhardware}. Also, for the given area, PIM has only half the number of MACs as compared to NDP due to area constraints determined by the DRAM technology \cite{UpMem,gomez2021benchmarking}. Thus, for ML applications involving latency-sensitive language and recommendation tasks, NDP has at least 10$\times$ speedup than other architectures.

\begin{figure}%
    \centering
    \subfloat[\centering Energy consumed by each layer\label{fig:FCLEnergy}]{{\includegraphics[width=0.75\linewidth]{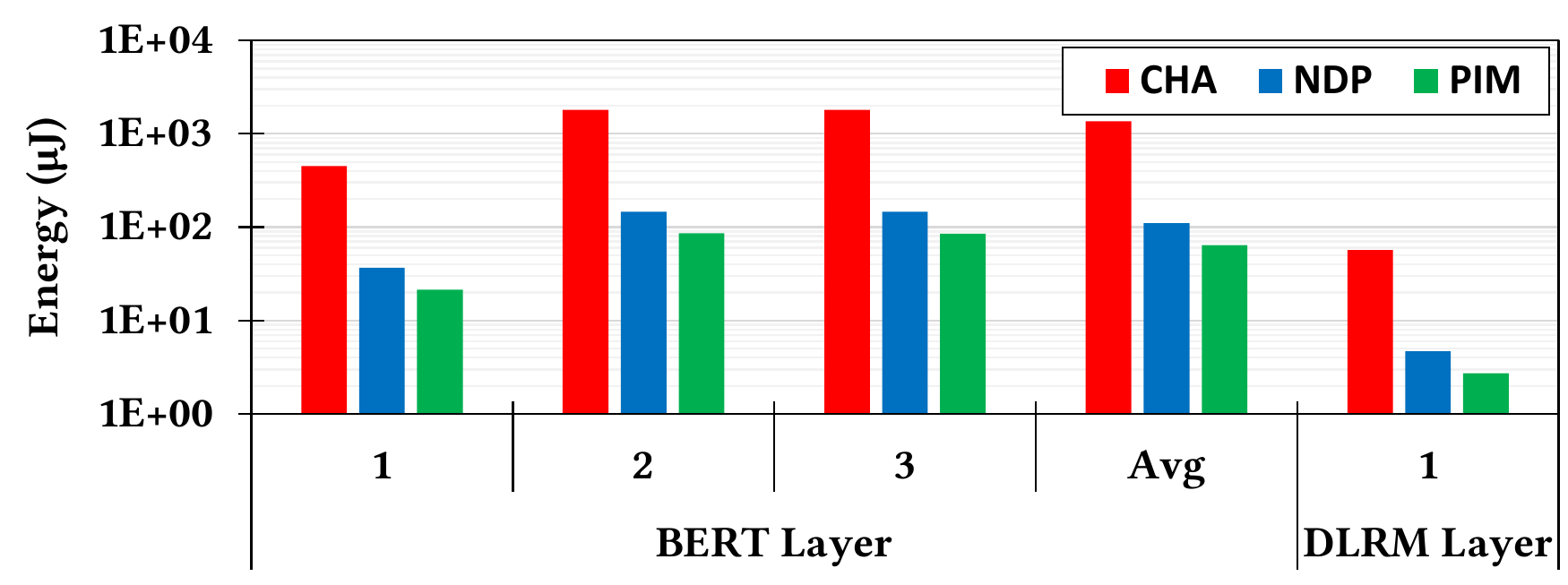} }}%
    \\
    \subfloat[\centering Breakup of energy per compute\label{fig:FCLEnergySplit}]{{\includegraphics[width=0.75\linewidth]{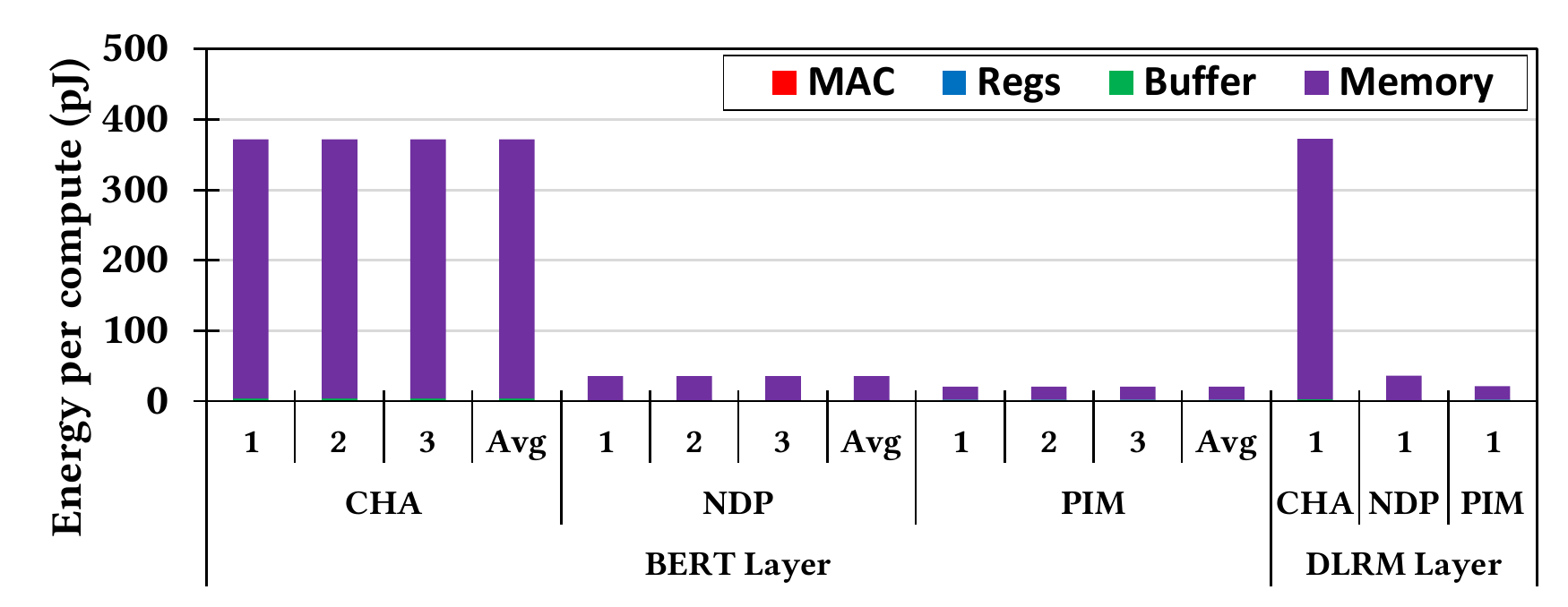} }}%
    \caption{Energy consumed in inferencing BERT and DLRM}%
    \label{fig:FCLEnergy2}%
\end{figure}

Fig.~\ref{fig:FCLEnergy} shows the total energy per layer for BERT and DLRM, and Fig.~\ref{fig:FCLEnergySplit} shows the split up of average energy per compute of BERT and DLRM.
PIM is the most efficient among the three architectures in terms of energy. PIM consumes 1.7$\times$ and 21$\times$ lesser energy as compared to NDP and CHA. CHA consumes 20$\times$ more energy to bring data from main memory to compute units than PIM during inference of FCL. This, along with the higher buffer energy, leads to the 21$\times$ higher energy expenditure in CHA compared to PIM. In PIM, the energy consumed by the MAC unit is twice that of NDP due to the DRAM process. However, the energy to access data from LLM is half compared to NDP. 
Thus, PIM is at least 40\% more energy efficient than other architectures for CNN and FCL ML workloads.

\section{Sensitivity Analysis Results}
 In this section we dive deeper into the individual components of each of these architectures to study their effect on the overall system. This analysis helps us to better understand the resource requirements for designing DNN accelerators. It will also help designers to identify components that need to be improved depending upon the requirements of the accelerator by reducing the design space. The results presented in this section are obtained by averaging the observed values of all the layer of respective workloads.

\subsection{Batching of workload for CNN}

\begin{figure}[htpb]
    \centering
    \subfloat[\centering Relative latency \label{fig:BatchingCNNTime}]{{\includegraphics[width=0.75\linewidth]{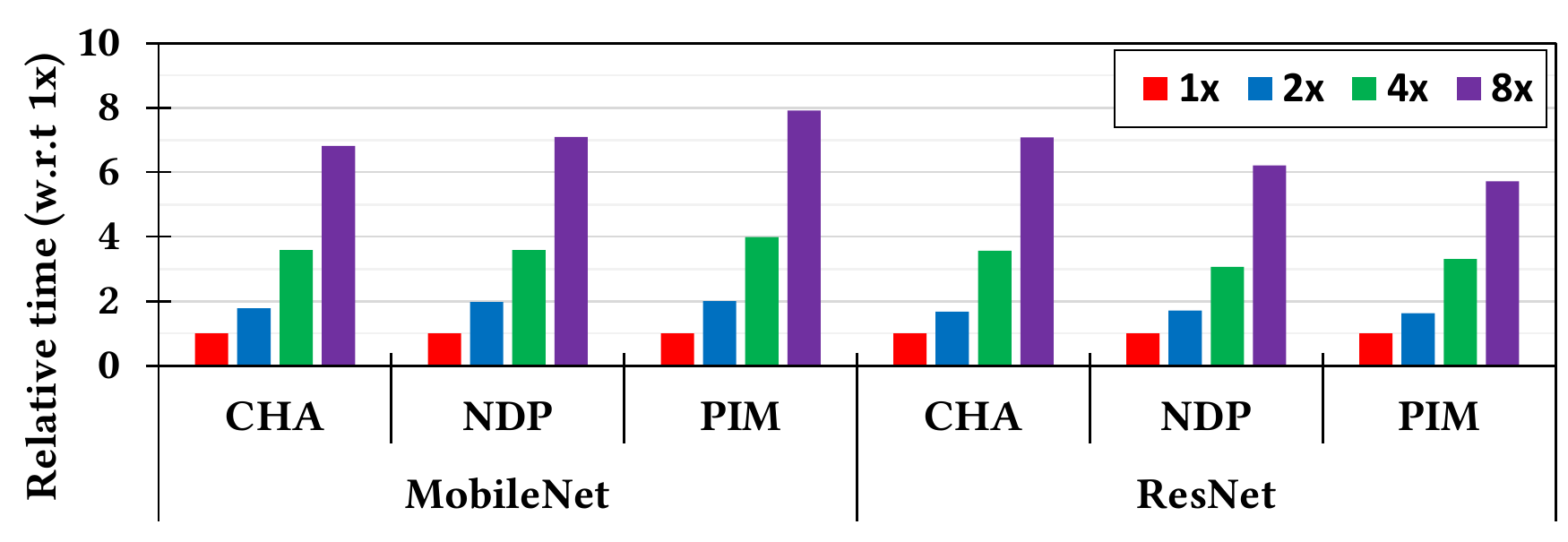} }}%
    \\
    \subfloat[\centering Relative utilization of MAC \label{fig:BatchingCNNUtilization}]{{\includegraphics[width=0.75\linewidth]{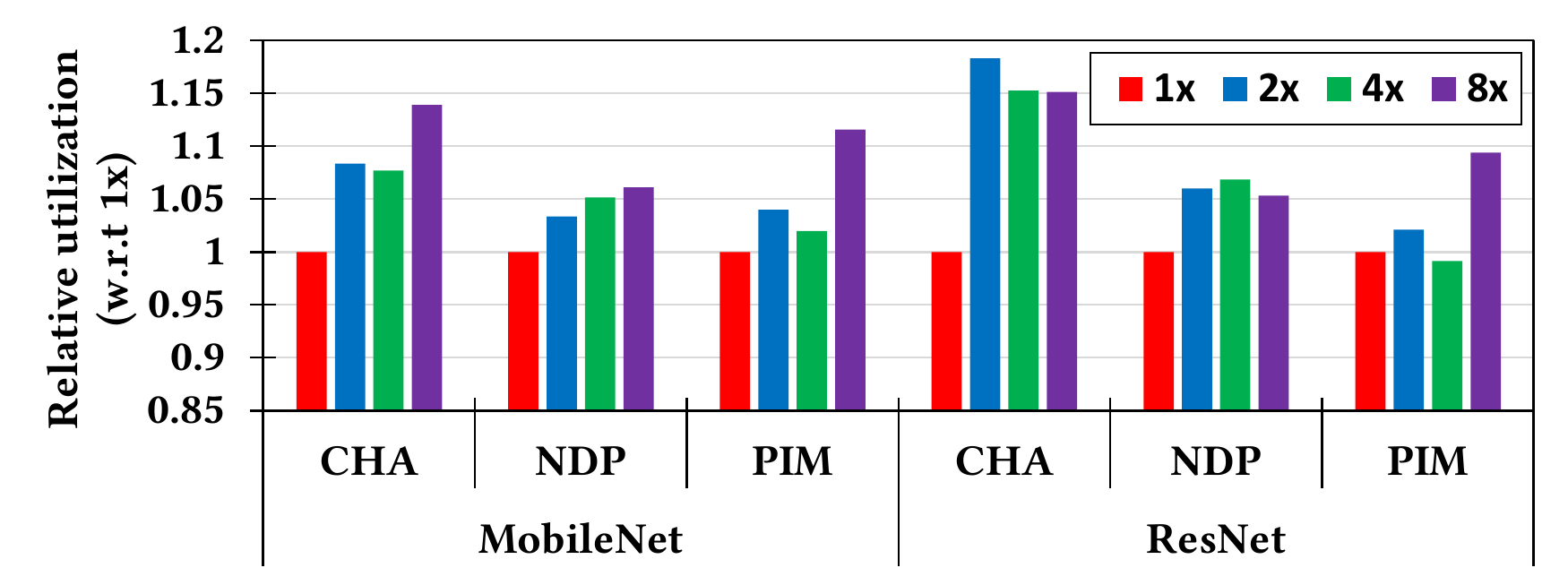} }}%
    \caption{Relative latency of MobileNet and ResNet for batch size of 1$\times$, 2$\times$, 4$\times$ and 8$\times$ on CHA, NDP and PIM}%
    \label{fig:BatchingCNNTime2}%
\end{figure}

Batching is a common technique used in many application scenarios within the ML domain where higher throughput is needed. When batched, multiple inputs are processed in parallel rather than sequentially. This parallel processing increases the average latency of processing each layer, however allows for the reuse of weights across multiple inputs. Batching becomes very important for the \emph{offline scenario} in MLPerf. \emph{Offline scenario} is for applications where all the data is readily available and there is no strict constraint on latency for single inference. Batching is also important for \emph{multi-stream} and \emph{server} applications where throughput is important with additional constraints on latency. Though batching increases throughput, the benefits are not uniform across architectures.

Fig.~\ref{fig:BatchingCNNTime2} shows relative latency for inference and relative MAC utilization of CHA, NDP and PIM for MobileNet and ResNet for batch sizes 1, 2, 4, and 8, respectively. 
We see that as the batch sizes increase by 2$\times$, 4$\times$, and 8$\times$, the latency on average, increases by 1.7$\times$, 3.5$\times$, and 6.8$\times$ as shown in Fig.~\ref{fig:BatchingCNNTime}. 
However, when doubling the batch size, throughput increases by 13\%, 8\% and 1\% for CHA, NDP and PIM, respectively, for MobileNet. Further, ResNet's throughput increases by 13\%, 25\% and 28\% for CHA, NDP and PIM, respectively, when batch size is doubled. This variation across paradigm and workload is due to the following reasons. (i) When examining the shape of layers of CNN workloads, we observe that in MobileNet, 60\% of the layers lead to an increase in the number of output channels, while in ResNet, only 30\% of the layers lead to an increase in the number of output channels, as shown in Table~\ref{tab:indextable}. Hence, ResNet is more symmetric across layers as compared to MobileNet, which leads to better resource utilization, and is, therefore, more suitable for batching, as shown in Fig.~\ref{fig:BatchingCNNUtilization}. (ii) The distributed architecture of NDP and PIM is not suitable for batching due to variations in sizes across layers. 

Further, we observed particular batch sizes work better than others for each paradigm. CHA and PIM have better throughput improvement at a batch size of 8 due to higher utilization of MACs for MobileNet and ResNet as MACs are spatially arranged with 8 MACs on one row and column, which leads to better parallelization. With its 4x4 arrangement, NDP has a better throughput for batch sizes of 4 and 8.

\begin{figure}[htpb]
    \centering
    \subfloat[\centering Relative energy\label{fig:CNNBatchingEnergy}]{{\includegraphics[width=0.75\linewidth]{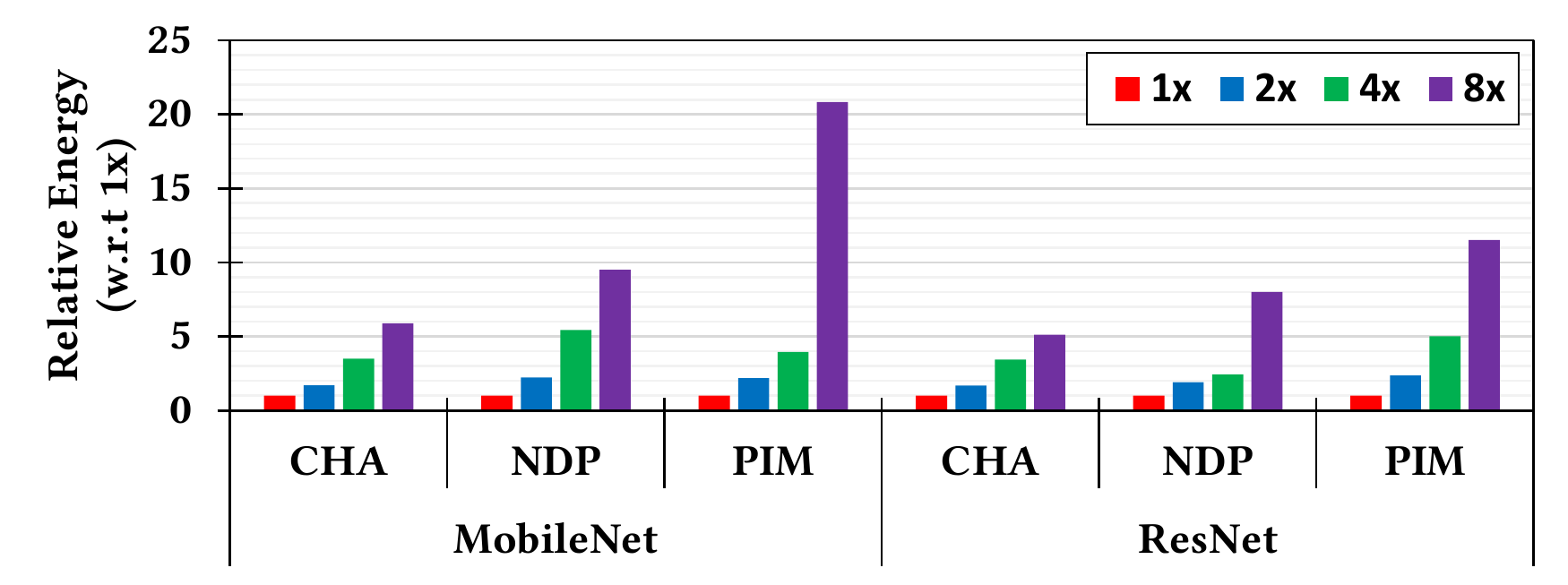} }}%
    \\
    \subfloat[\centering Breakup of energy per compute \label{fig:BatchingCNNEnergyPerCompute}]{{\includegraphics[width=0.75\linewidth]{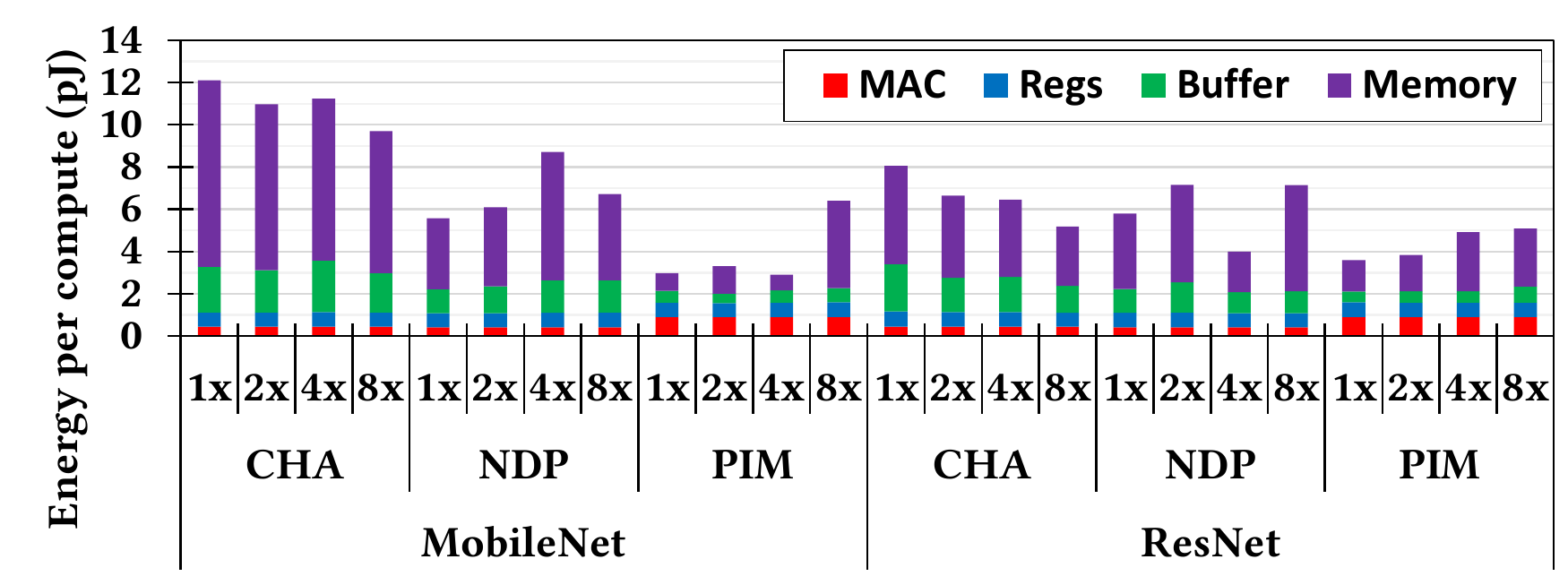} }}%
    \caption{Energy consumed for MobileNet and ResNet for batch size of 1$\times$, 2$\times$, 4$\times$ and 8x}%
    \label{fig:CNNBatchingEnergy2}%
\end{figure}

Fig.~\ref{fig:CNNBatchingEnergy} shows the relative energy when batching MobileNet and ResNet workloads on the accelerators. Further, Fig.~\ref{fig:BatchingCNNEnergyPerCompute} shows the split of energy per compute for Fig.~\ref{fig:CNNBatchingEnergy}.
For MobileNet using CHA, batching causes reduction in energy per layer by 10\%, 10\%, and 20\% for 2$\times$, 4$\times$, and 8$\times$ batching respectively, compared to 12$\times$.
For ResNet using CHA, it is 15\%, 15\%, and 35\% for 2$\times$, 4$\times$, and 8$\times$ batching respectively. The decrease in energy is attributed to decreased last-level memory accesses in CHA due to data reuse. 

Contrary to CHA, in NDP and PIM, batching increases the number of memory accesses leading to increase in energy consumption. In the worst case for PIM, the energy consumption can increase upto 2.5$\times$. Hence, for CNN, batching is beneficial for CHA, increasing throughput and decreasing overall energy. But NDP and PIM need to be redesigned with better support for batching as the current buffer to PE ratio leads to larger number of memory access with increase in batch sizes.

\subsection{Batching of workload for FCL}
\begin{figure}[htpb]
    \centering
    \subfloat[\centering Relative latency\label{fig:BatchingFCLTime}]{{\includegraphics[width=0.75\linewidth]{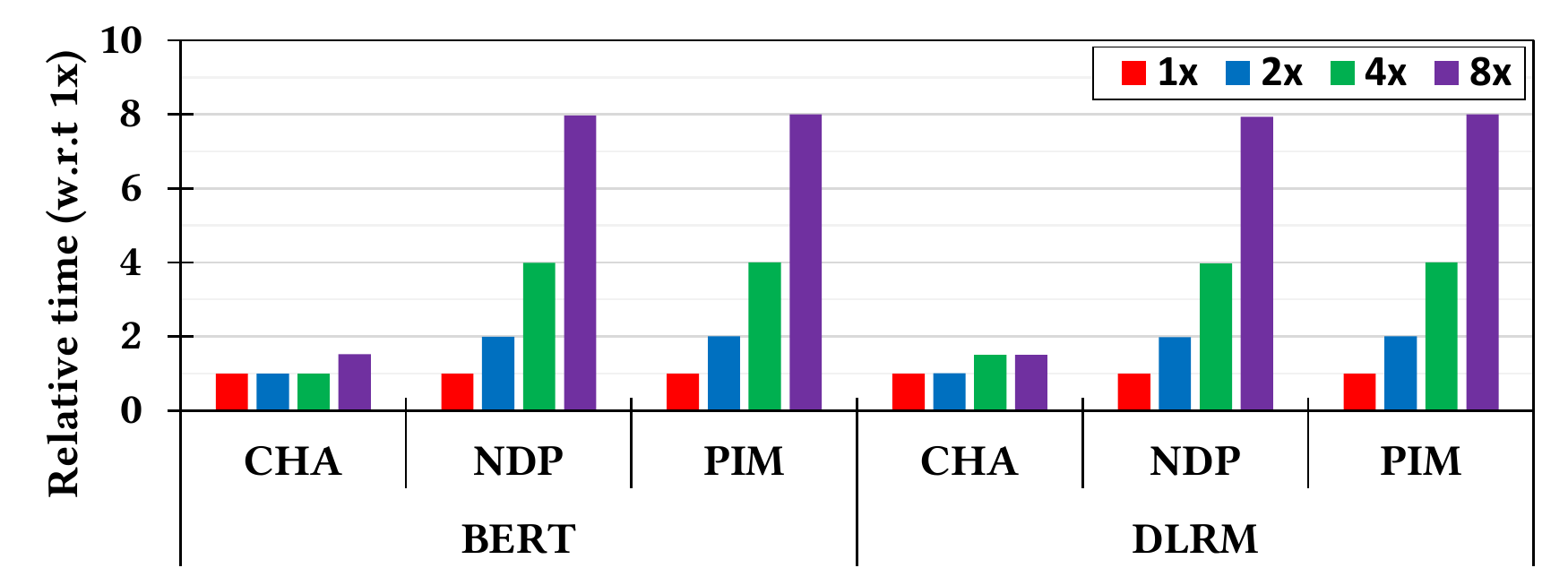} }}%
    \\
    \subfloat[\centering Relative utilization of MAC\label{fig:BatchingFCLUtilization}]{{\includegraphics[width=0.75\linewidth]{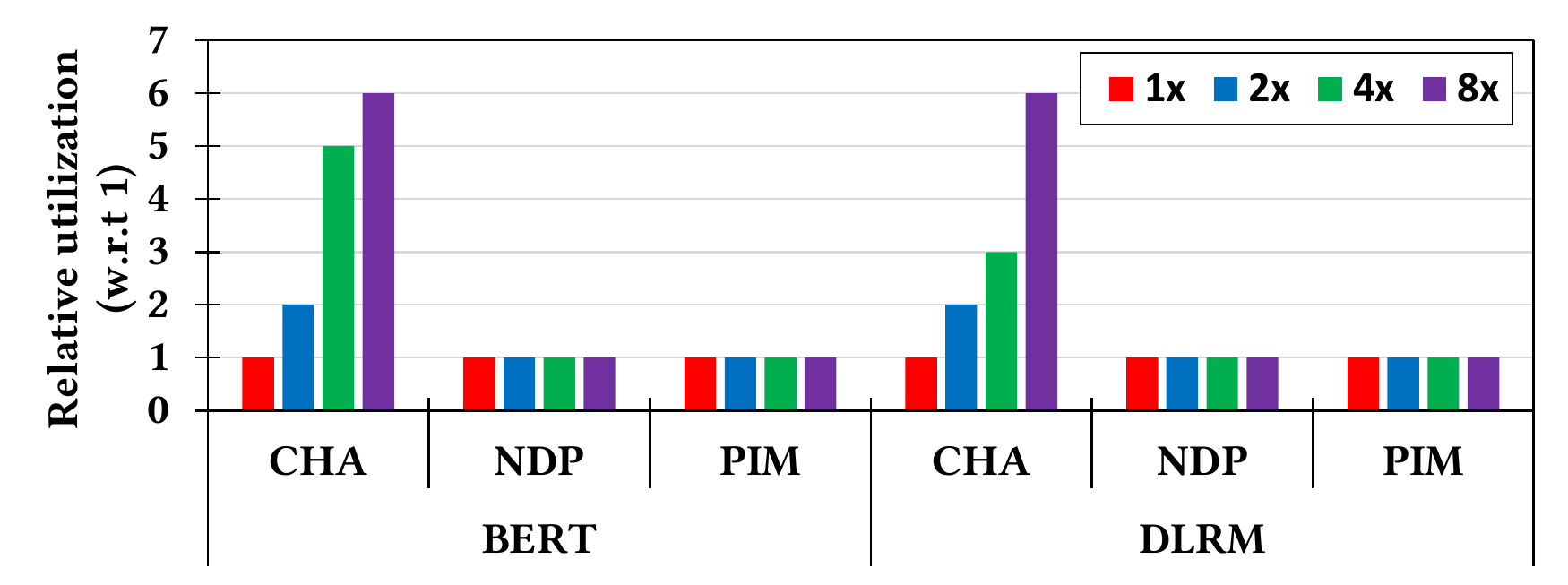} }}%
    \caption{Relative latency and MAC utilization of BERT and DRLM for batch size of 1$\times$, 2$\times$, 4$\times$ and 8$\times$}%
    \label{fig:BatchingFCLTime2}%
\end{figure}

When FCL workloads are batched, inputs are processed in parallel, compared to processing them one after the other. In this parallel processing, the weight matrix is reused for each vector input.
Figures ~\ref{fig:BatchingFCLTime} and ~\ref{fig:BatchingFCLUtilization} show the average latency for inference per layer and the relative MAC utilization, respectively, for BERT and DLRM when batching at 1$\times$, 2$\times$, 4$\times$, and 8$\times$.
For FCL layers, there is no improvement in throughput for NDP and PIM as they are already at 100\% utilization at 1$\times$, as seen in Fig.~\ref{fig:FCLUtilization}. But for CHA, on average, the throughput and utilization increase by 2$\times$, 3.3$\times$, and 5.3$\times$ for a batch size of 2$\times$, 4$\times$, and 8$\times$. This is because NDP and PIM have no last-level memory bandwidth bottleneck, whereas CHA is limited by LLM bandwidth bottleneck for FCL.
Thus, as data reuse potential increases with batching, CHA can reuse the data across multiple MACs while keeping the number of memory accesses the same, increasing throughput. 

\begin{figure}[h]
    \centering
    \subfloat[\centering Relative energy\label{fig:BatchingFCLEnergy}]{{\includegraphics[width=0.75\linewidth]{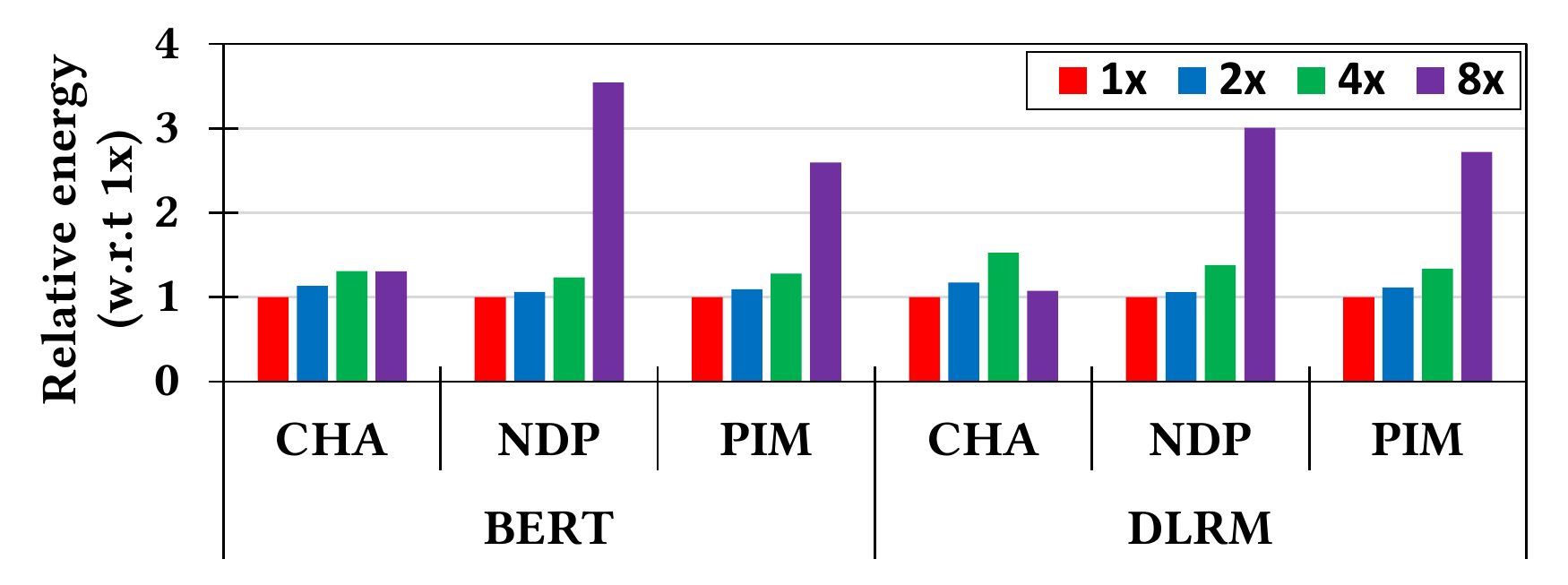} }}%
    \\
    \subfloat[\centering Breakup of energy per compute\label{fig:BatchingFCLEnergyPerCompute}]{{\includegraphics[width=0.75\linewidth]{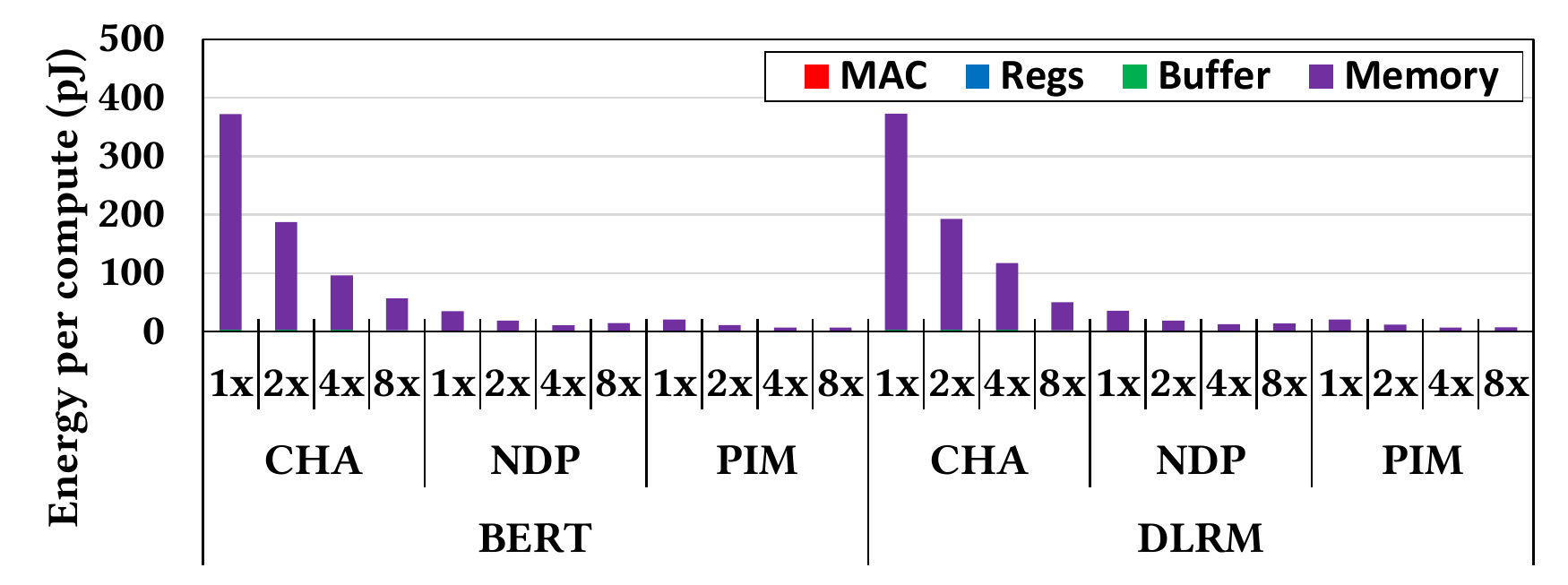} }}%
    \caption{Energy consumed for BERT and DLRM for batch size of 1$\times$, 2$\times$, 4$\times$ and 8x}%
    \label{fig:BatchingFCLEnergy2}%
\end{figure}

Figures ~\ref{fig:BatchingFCLEnergy}~and~\ref{fig:BatchingFCLEnergyPerCompute} show the relative energy and split up of energy per compute when batching BERT and DLRM workloads on the accelerators.
Batching reduces energy requirement per layer by 0.8$\times$ and 2$\times$ at a batch size of 2$\times$ and 4$\times$ compared to processing them sequentially for all the architectures. This reduction is due to the amortization of memory access energy across the layers.
However, for CHA, at 8$\times$ batching, the energy consumption per layer is 7$\times$ lower than 1$\times$ batch size, but for NDP and PIM, the energy consumption per layer is only 2.7$\times$ lower compared to 1$\times$ batch size. This is because the optimal mapping for latency bypasses the global buffer in NDP and PIM. At the same time, the registers in the PE are not able to hold all the input and output words till all the operations associated with those words are finished leading to an increase in the number of LLM access. 
Hence for FCL layers, it is beneficial to batch the inputs for all three accelerators as it considerably decreases energy consumption. 

\subsection{Last-level-memory bandwidth for CNN and FCL}

\begin{figure}[htpb]
    \centering
    \subfloat[\centering Relative latency \label{fig:ExternalBWCNNTime}]{{\includegraphics[width=0.75\linewidth]{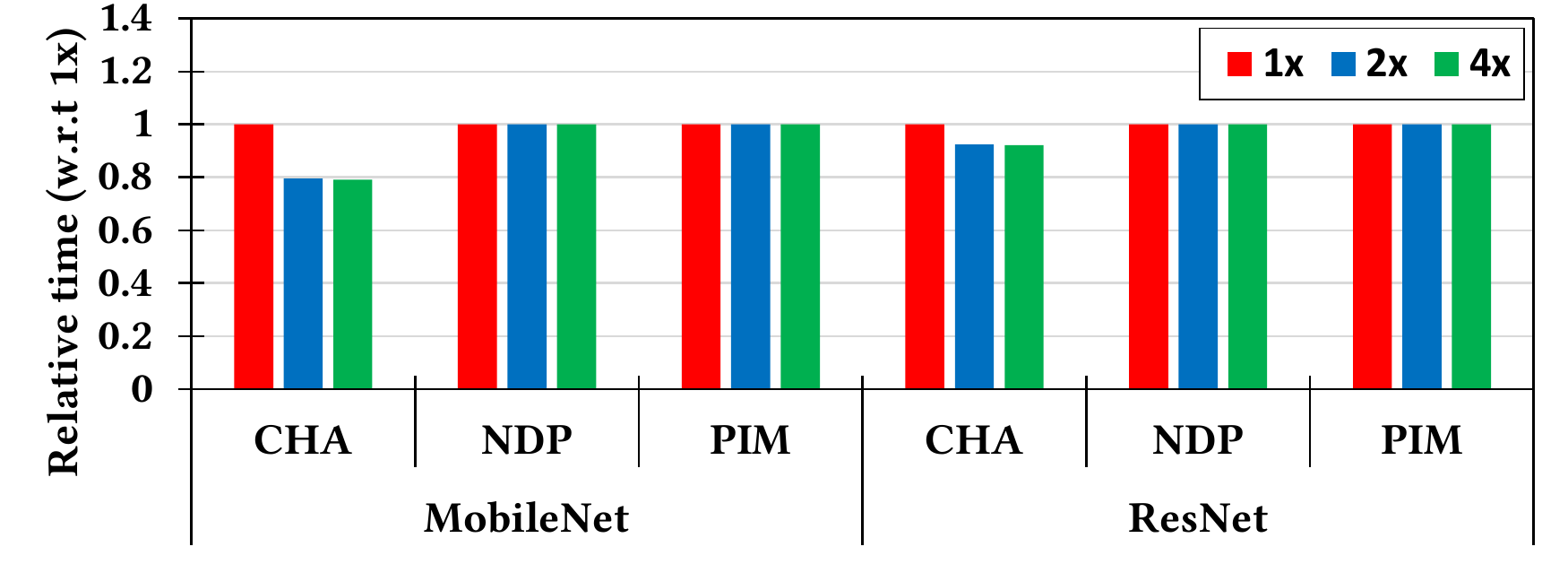} }}%
    \\
    \subfloat[\centering Relative utilization of MAC \label{fig:ExternalBWCNNUtilization}]{{\includegraphics[width=0.75\linewidth]{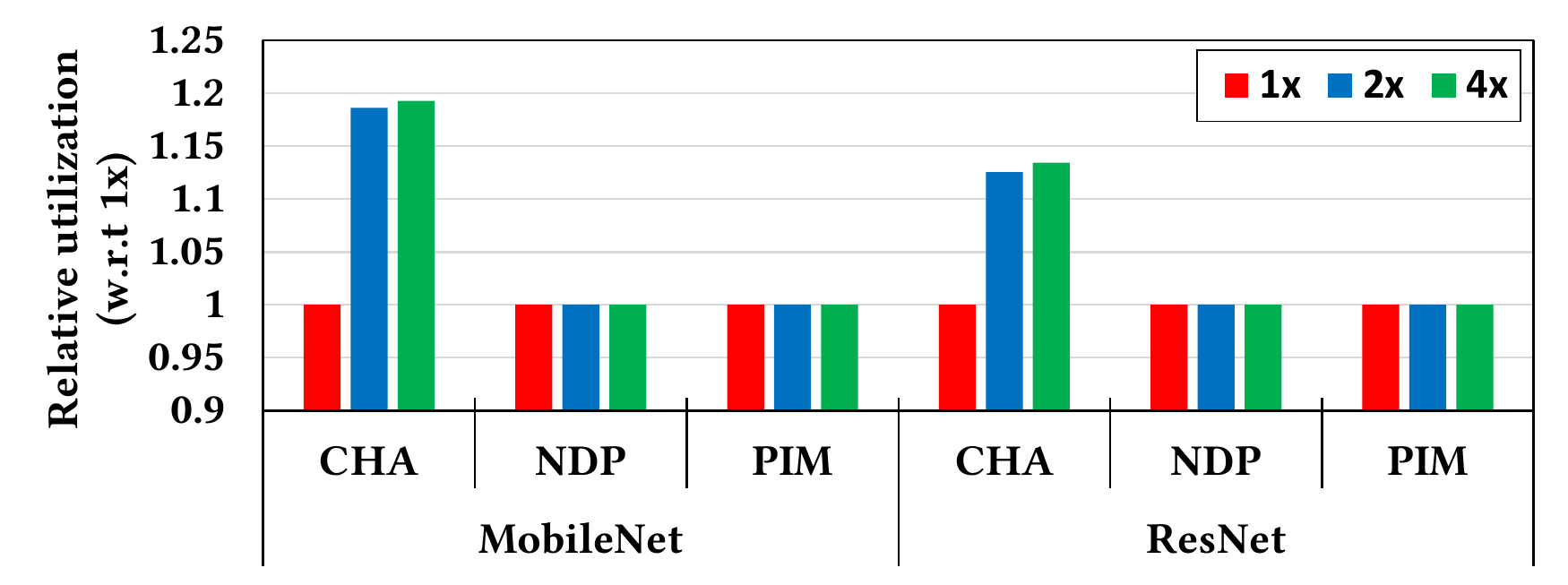} }}%
    \caption{Latency and MAC utilization of MobileNet and ResNet for LLM bandwidth = 1$\times$, 2$\times$ and 4x}%
    \label{fig:ExternalBWCNNTime2}%
\end{figure}

Figures ~\ref{fig:ExternalBWCNNTime} and ~\ref{fig:ExternalBWCNNUtilization} show the relative latency and MAC utilization for MobileNet and ResNet workloads with respect to a 2$\times$ and 4$\times$ increase in the LLM bandwidth, respectively. The baseline LLM bandwidth of each accelerator paradigm is given by the \emph{DRAM-Compute bandwidth} entry in Table \ref{tab:constraints}.
For NDP and PIM, increasing the LLM bandwidth does not improve performance as their performance is not limited by memory bandwidth. For CHA, latency decreases by 20\% for MobileNet and 10\% for ResNet when the bandwidth is doubled. CHA is able to provide more data for MAC units to process, leading to an increase in MAC utilization and hence performance. However, at 4$\times$ bandwidth, no significant improvement is noted as CHA becomes compute bottlenecked.  Further, as MobileNet has lower data reuse potential, as shown in Table \ref{tab:indextable}, higher bandwidth provides more benefits to MobileNet as compared to ResNet.

\begin{figure}[htpb]
    \centering
    \subfloat[\centering Relative latency\label{fig:ExternalBWFCLTime}]{{\includegraphics[width=0.75\linewidth]{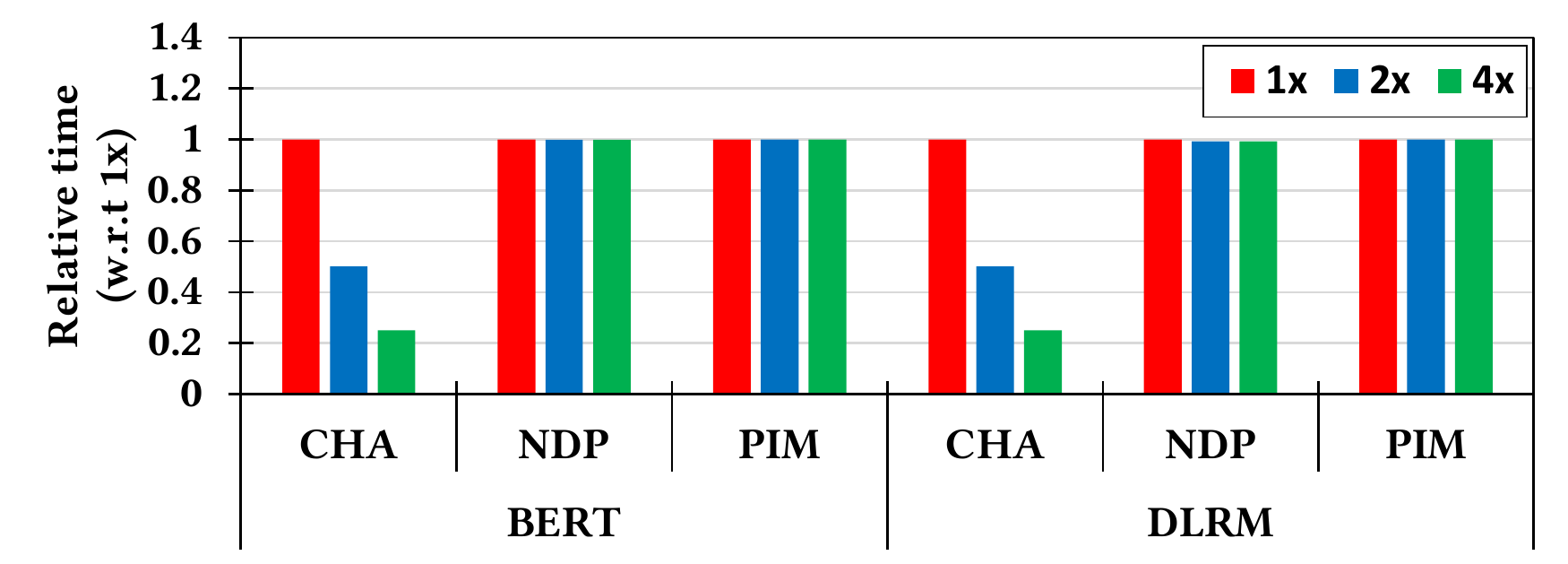} }}%
    \\
    \subfloat[\centering Relative utilization of MAC \label{fig:ExternalBWFCLUtilization}]{{\includegraphics[width=0.75\linewidth]{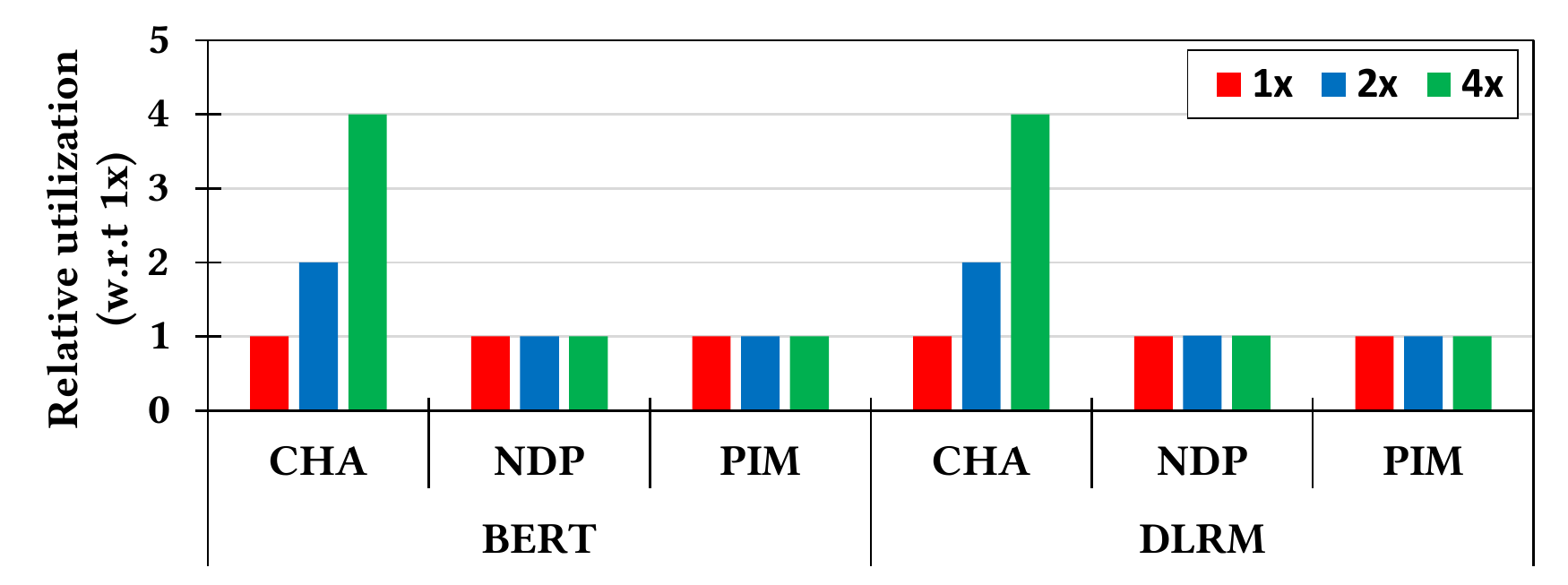} }}%
    \caption{Latency and MAC utilization of BERT and DLRM for LLM bandwidth = 1$\times$, 2$\times$ and 4x}%
    \label{fig:ExternalBWFCLTime2}%
\end{figure}

Figures ~\ref{fig:ExternalBWFCLTime} and ~\ref{fig:ExternalBWFCLUtilization} shows the average relative latency per layer and MAC utilization for BERT and DLRM workloads, respectively, when LLM bandwidth is increased.
For NDP and PIM, increasing the LLM bandwidth does not improve performance similar to CNN. 
However, as data reuse potential is low for FCL layers, as shown in Table \ref{tab:indextable}, increasing LLM bandwidth greatly improves the latency of CHA for FCL. Increasing the bandwidth by 2$\times$ and 4$\times$ increases utilization by 2$\times$ and 4$\times$, leading to a speedup of 2$\times$ and 4$\times$, respectively. 

Overall, NDP and PIM do not benefit from the increase in LLM bandwidth for both CNN and FCL workloads. Since CHAs are limited by memory bandwidth for both CNN and FCL, we see improvement in performance with an increase in bandwidth. In FCL, we see more benefits as compared to CNN for CHA as data reuse is lower in FCL.

\subsection{Working memory size} \label{sensitibuffersize}

\begin{figure}%
    \centering
    \subfloat[\centering Latency\label{fig:BufferSizeCNN 
    Time}]{{\includegraphics[width=0.75\linewidth]{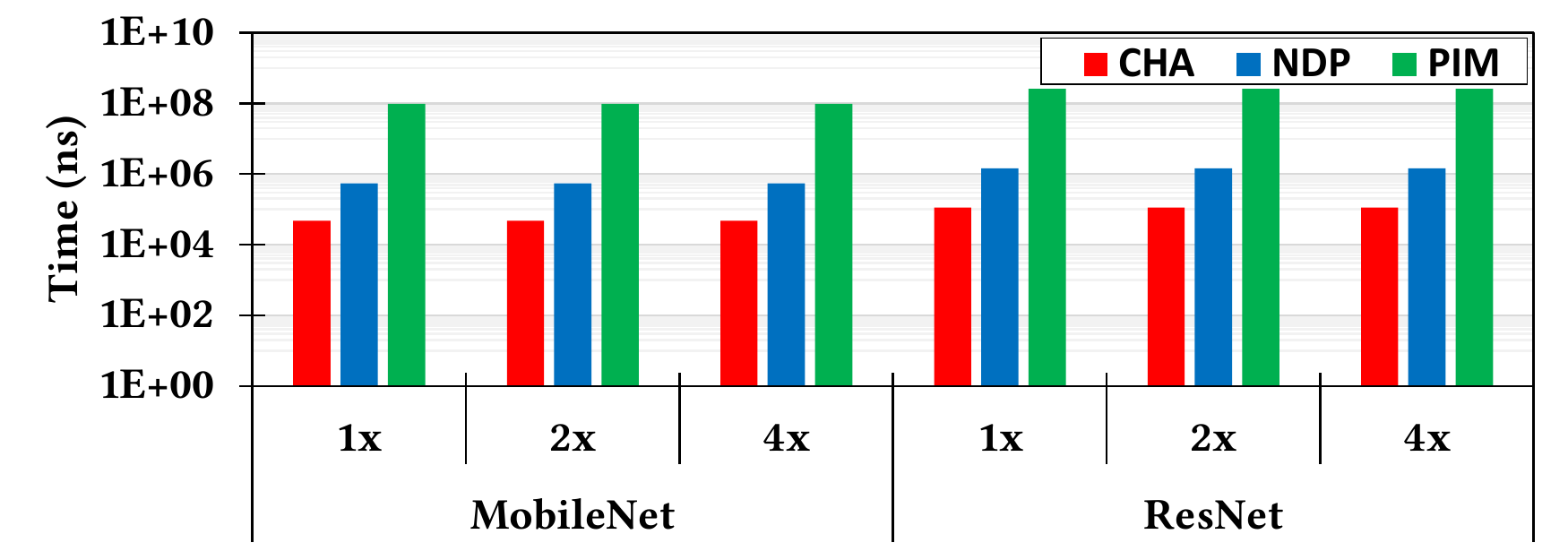} }}%
    \\
    \subfloat[\centering Energy per Compute \label{fig:BufferSizeCNN}]{{\includegraphics[width=0.75\linewidth]{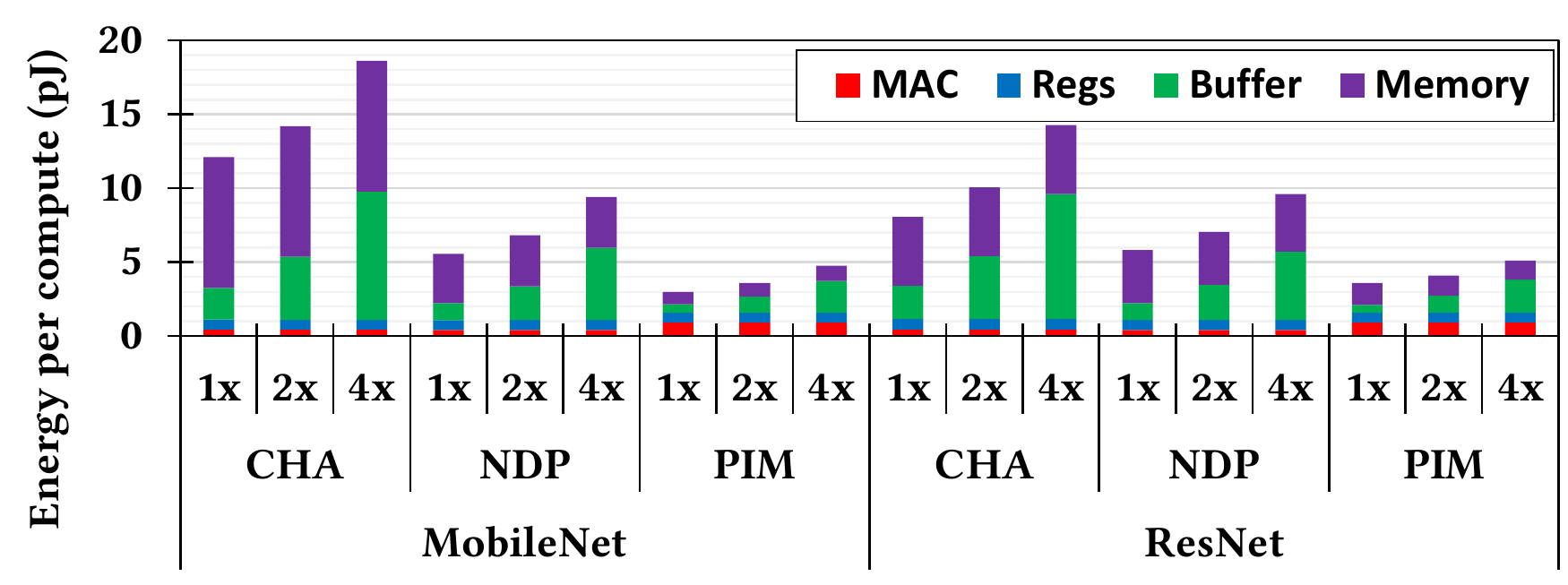} }}%
    \caption{Latency and Energy of CNN workloads when Global Buffer is increased 1$\times$, 2$\times$ and 4x}%
    \label{fig:BufferSize}%
\end{figure}

Fig.~\ref{fig:BufferSizeCNN} shows the split of energy per compute for CNN workloads when working memory size is increased 2$\times$ and 4$\times$, compared to baseline designs. In CHA and NDP, global buffer represents the working memory, as seen in Fig. \ref{fig:Architectures}. The working memory of PIM is shown in Fig. \ref{fig:NHAUpMem}.
It is observed that increasing working memory size of CHA, NDP, and PIM leads to a negligible change in latency and utilization of MAC, as shown in Fig.~\ref{fig:BufferLayoutTime}. However, as the access energy of the buffers has increased, the energy per compute increases. Hence, minimizing the buffer size to the required application will reduce energy requirement.

\subsection{Buffer layout}
\begin{figure}[htpb]%
    \centering
    \subfloat[\centering Latency]{{\includegraphics[width=0.75\linewidth]{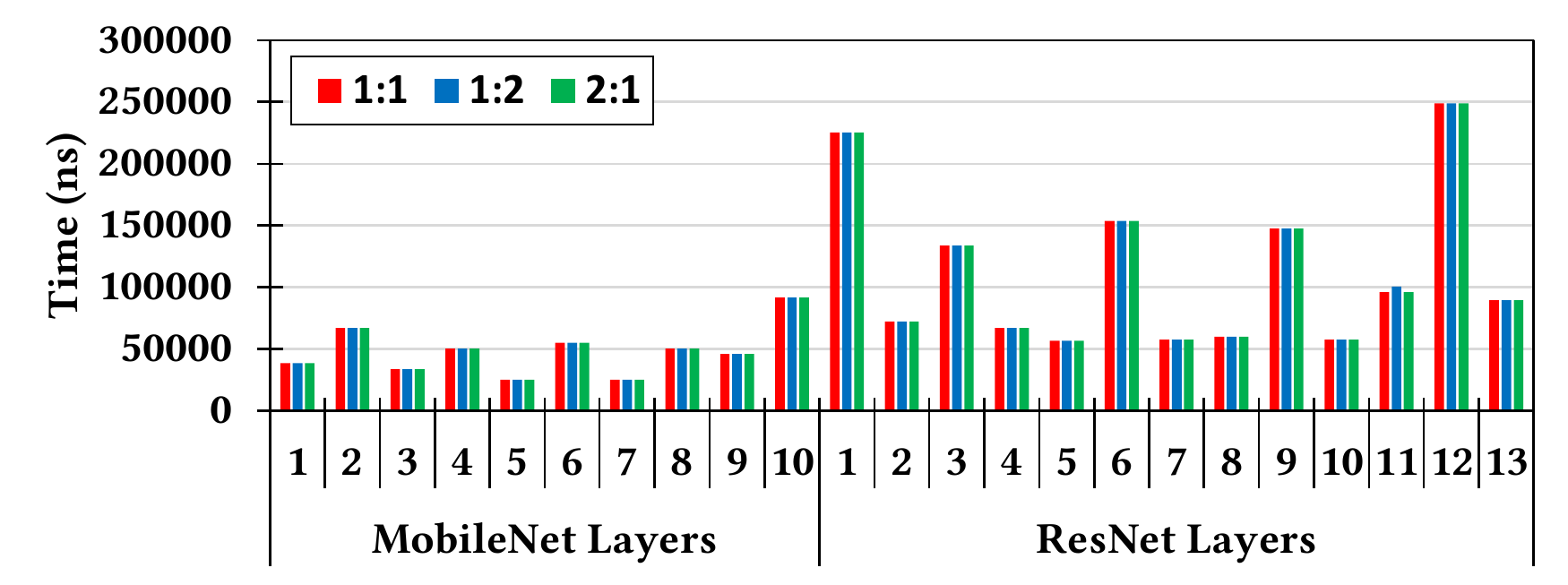} }}%
    \\
    \subfloat[\centering Energy]{{\includegraphics[width=0.75\linewidth]{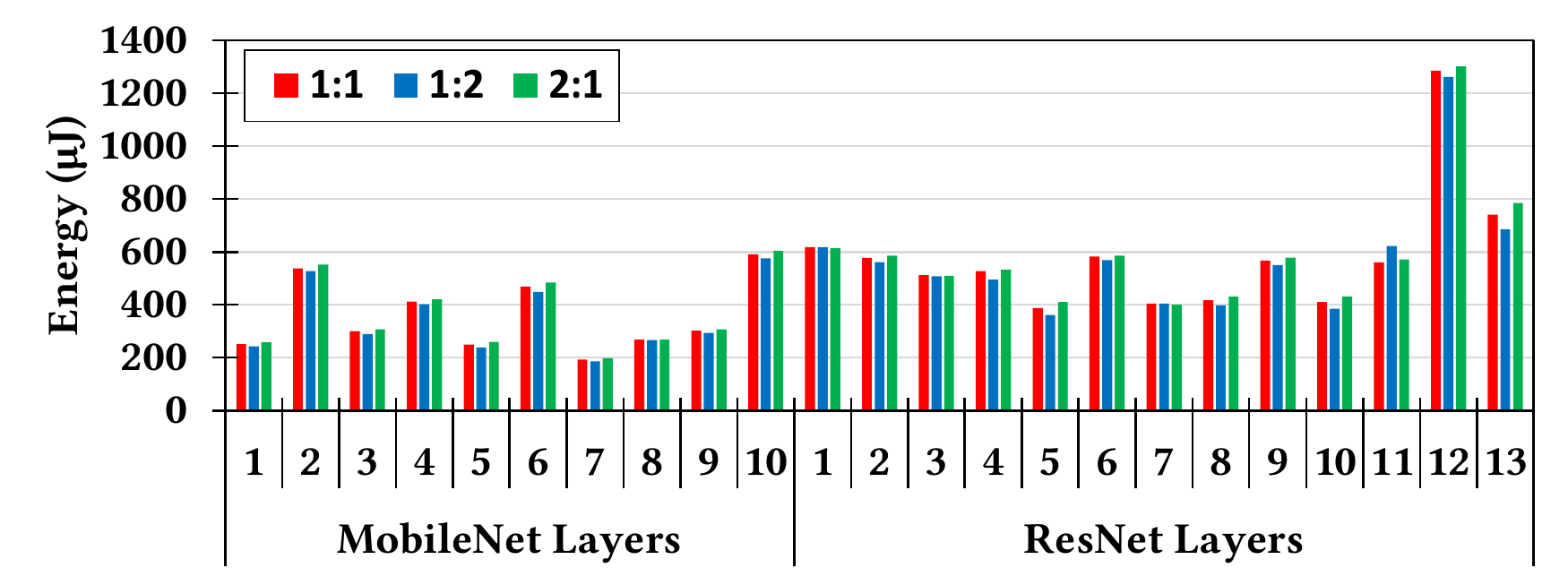} }}%
  
    \caption{Latency and energy of MobileNet and BERT on CHA when 3MB of buffer is distributed between global buffer and local buffer}%
    \label{fig:BufferLayoutTime}%
\end{figure}

\begin{figure}[htpb]
  \centering
  \includegraphics[width=\linewidth]{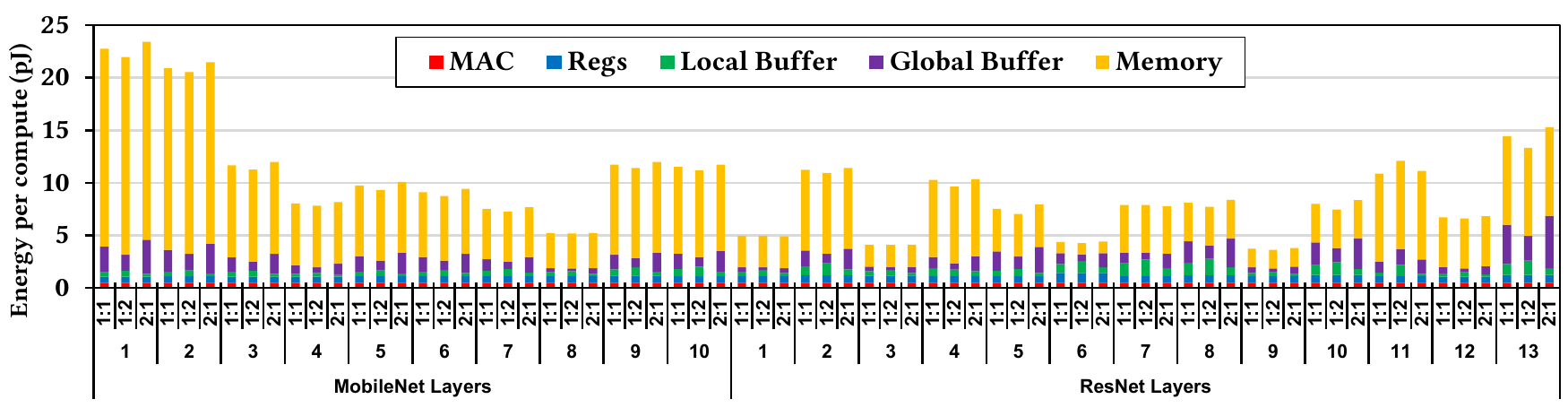}
  \caption{Per compute energy of MobileNet and BERT on CHA when 3MB of buffer is distributed between global buffer and local buffer}
 
  \label{fig:BufferLayoutPercompute}
\end{figure}

CHAs usually have at least two levels of buffers, one at the global level and one at the PE level (local buffer). The global buffer helps establish chip-level stationary approaches, and local buffers help in partitioning the workload. Some of the workload data is replicated across the PEs based on mapping to facilitate PE level computation.
In this analysis, the total buffer size is 3MB, and we partition it across global and local buffers. 
Fig.~\ref{fig:BufferLayoutPercompute} show energy per compute of MobileNet and ResNet, respectivley, on CHA with different buffer partion ratios.
For example, in \emph{1:1 configuration} 1.5 MB is allocated to both the global buffer and the local buffer distributed across 16 PEs.

We observe that the latency, as shown in Fig. \ref{fig:BufferLayoutTime}, remains same for all configurations except for layer 11 of ResNet. It is observed that the energy value is highest for 2:1 followed by 1:1 and lowest for 1:2 configurations.
These changes are due to resultant optimal mapping of the layer to PEs and buffers, reducing the number of reads and writes to the global buffer where the read and write energy is higher.

Fig.~\ref{fig:BufferLayoutPercompute} shows that the total buffer energy per compute reduces by 10$\%$ and 23$\%$ for 1:1 and 1:2 configurations compared to 2:1 configurations.
This is because the increase in access energy of local buffer compared to the decrease in access energy of global buffer is much smaller. However, the global buffer size to local buffer size ratio can not be skewed too much, as seen in layer 11 of ResNet. In this layer, there is diminished energy saving for 1:2 configuration compared to other configurations due to more accesses to the local buffer.

\subsection{Limits of performance for architecture paradigms.}

\begin{figure}[htpb]

    \centering
    \subfloat[\centering MobileNet and ResNet\label{fig:maxMACCNN}]{{\includegraphics[width=0.75\linewidth]{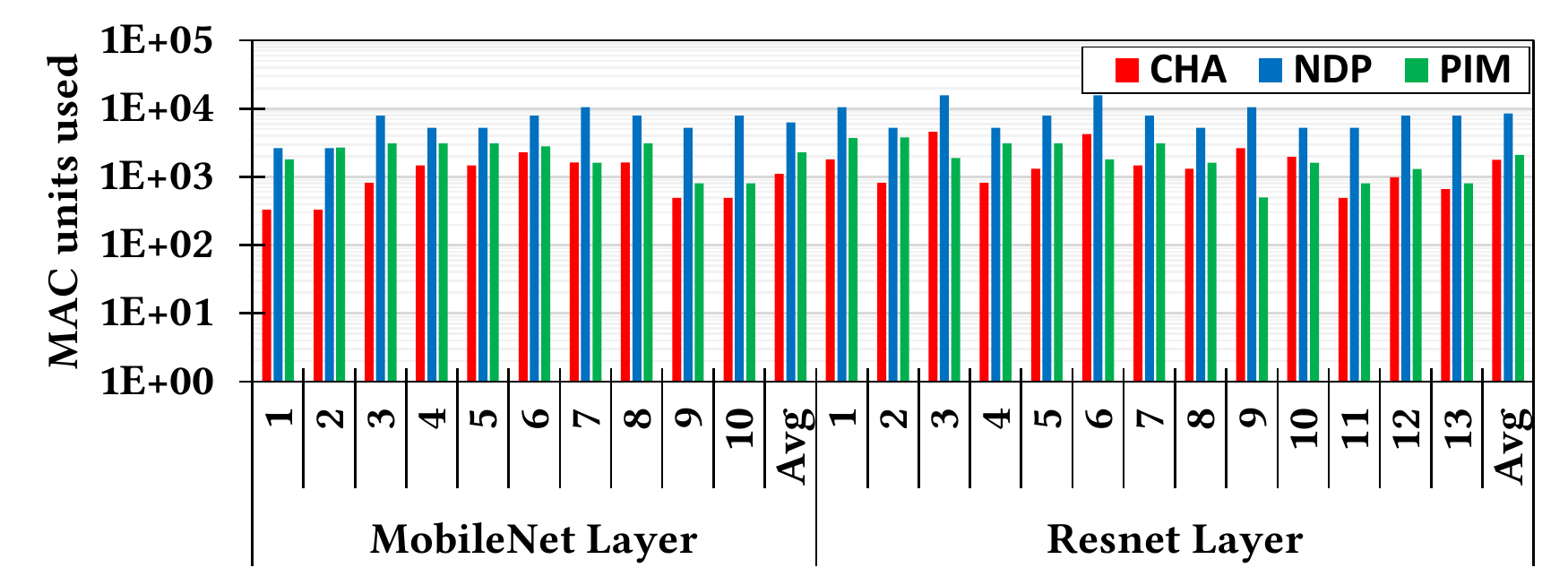} }}%
    \\
    \subfloat[\centering BERT and DLRM\label{fig:maxMACFCL}]{{\includegraphics[width=0.75\linewidth]{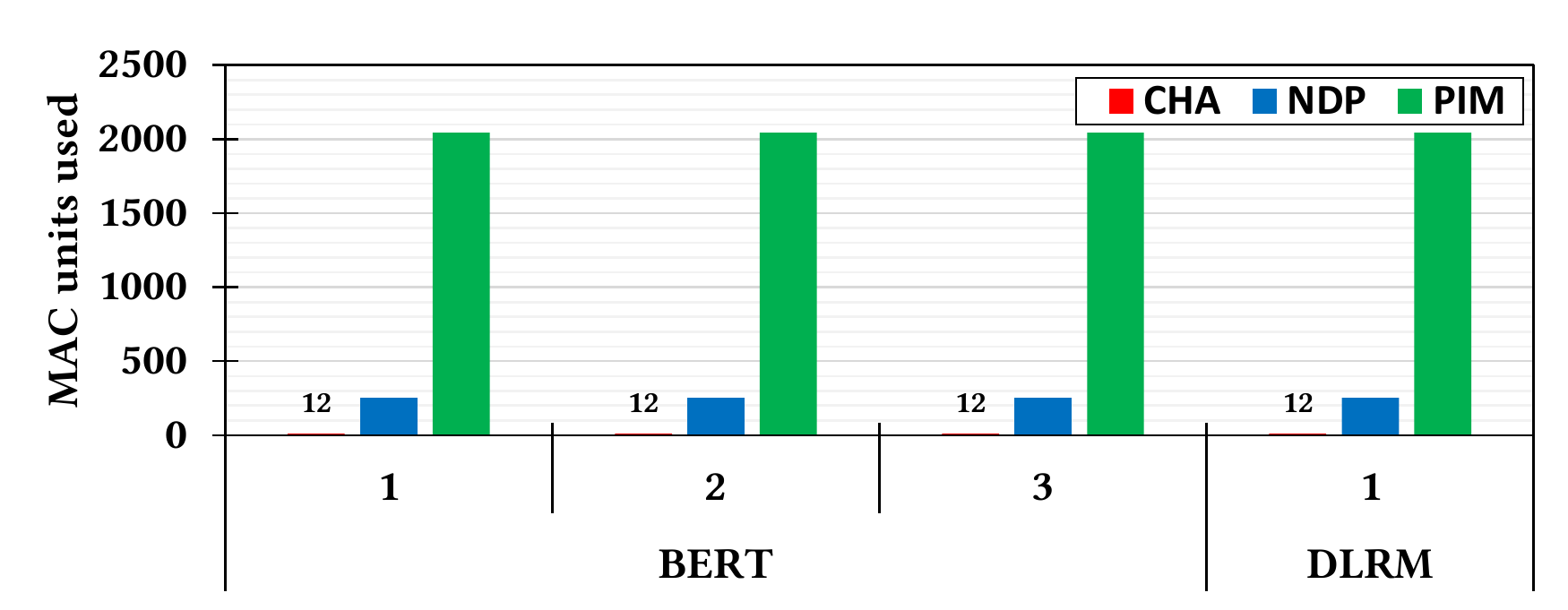} }}%

    \caption{Maximum MAC units used per layer of CNN and FCL workloads}%
    \label{fig:maxMAC2}%
  
\end{figure}

In this experiment, we relax the area and thermal constraints by increasing the resources available within the DNN accelerator models to gain insight into the impact of resources on various metrics. The LLM bandwidth is kept the same as the baseline design.
 To identify the maximum number of MACs that can be used per layer, the number of PEs are increased from the baseline such that the total number of MAC unit is 100 thousand in the MAX variants of CHA, NDP, and PIM.
Fig.~\ref{fig:maxMACCNN} shows the maximum MAC units used per layer of MobileNet and ResNet on CHA, NDP, and PIM, while Fig.~\ref{fig:maxMACFCL} shows the same data for BERT and DLRM workloads.
For MobileNet, half of the layers can use more than 1024 MACs (Baseline), while in ResNet, only 8 out of the 13 layers can use more than baseline. 
For NDP and PIM, all CNN layers can use more MACs than their respective baselines. 
The average MACs required per layer for MobileNet during inference are 1100, 6300, 2300 for CHA, NDP, and PIM, respectively. Due to its higher data reuse potential, ResNet has 1777, 8500, and 2100 average MACs per layer requirements for CHA, NDP, and PIM, respectively. 
This shows that NDP and PIM are extremely compute bottlenecked for CNN-based ML applications, and improving the number of MACs can potentially improve the performance of NDP and PIM. 

For FCL layers, with optimal mapping, the total count of MAC units that can be used is limited to 12, 256 and 2100 with the available LLM bandwidth. Hence, for FCL layers, CHA is not compute bottlenecked but is severly LLM bandwidth limited. In NDP, the LLM bandwidth balances the compute requirement in the baseline design for FCL workloads. Hence, additional PEs in the MAX variant are not used. An increase in the average number of MAC units used per layer in the MAX variant of PIM shows that baseline PIM is compute bottlenecked, and future designs should have 16$\times$ more MACs in the PIM chip to balance compute and bandwidth for FCL workloads.

\begin{figure}[htpb]

  \centering
  \includegraphics[width=\linewidth]{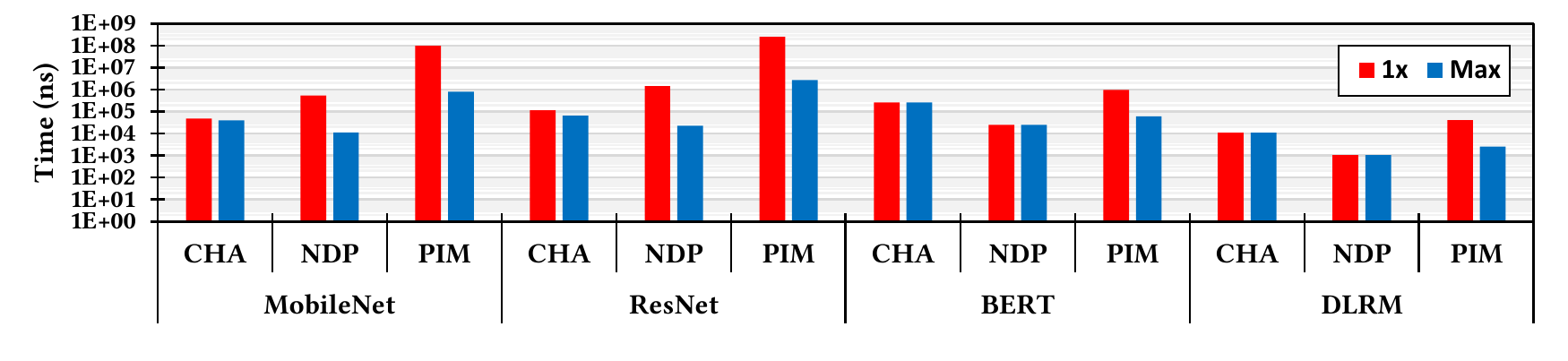}

  \caption{Comparison of time taken to compute by baseline systems against corresponding MAX variant systems with 100 thousand MAC units for MobileNet, ResNet, BERT and DRLM}
 
  \label{fig:maxMACTime}
\end{figure}

Further, we consider the resultant latency if all compute bottleneck is removed. 
Fig.~\ref{fig:maxMACTime} shows the latency of baseline design and the MAX variant with the MAC count increased to 100 thousand. There are several key observations. (i)~There is no significant increase in performance for MAX CHA as compared to baseline CHA for both CNN and FCL workloads due to LLM bandwidth bottleneck. (ii)~For CNN workloads, there is a 50$\times$ speedup for MAX NDP due to high data reuse. Whereas in FCL workloads, there is no latency improvement due to no data reuse. (iii)~MAX NDP has 3$\times$ lower latency than MAX CHA, making it the fastest across both workloads. (iv)~Compared to baseline PIM, MAX PIM has 120$\times$ and 15$\times$ speedup for CNN and FCL workloads, respectively. The significant difference in speedup is due to higher MAC utilization in MAX PIM in CNN because of data reuse. (v)~We also observe that MAX PIM has lower latency per layer than MAX CHA for FCL workloads, hence could be a better paradigm for FCL workloads. However, MAX PIM is not better than baseline CHA for CNN workloads. 

\section{Inferences from the Experiments}

In this section, we summarize the significant qualitative findings from the detailed comparison of representative architectures from the three DNN accelerator paradigms when inferencing CNN and FCL workloads. We observe the following. (i) CHA has the least latency per layer for CNN workloads. (ii) NDP has the least latency per layer for FCL workloads. (iii) PIM consumes the least energy for CNN and FCL workloads.  

Further, in order to identify how latency and energy characteristics can be improved for each workload on these representative architectures, we performed sensitivity analysis on significant components of the accelerator and found the following. (i) Batching of CNN and FCL workloads can decrease inference energy on CHA, NDP, and PIM. However, NDP and PIM should be redesigned to support larger batch sizes. Otherwise, this could lead to higher energy with larger batch sizes. (ii) Batching of FCL workloads in CHA leads to a linear increase in throughput at a negligible increase in latency of processing each layer as it heavily benefits from data reuse. (iii) Batching FCL workloads on NDP and PIM can be detrimental for latency-sensitive applications because with the increase in batching the latency of the network also increases. (iv) Increasing LLM bandwidth for CHA can linearly decrease latency for FCL applications and would be highly beneficial for latency bound single-stream applications. (v) Total on-chip buffer can be redistributed, based on workload, between the different buffer hierarchies in CHA to reduce on-chip energy. Further, an increase in on-chip buffer size beyond data reuse requirements leads to higher on-chip energy with no additional benefits.

Additionally, we identified the limits of each paradigm for CNN and FCL workloads to identify future possibilities and found the following. (i) NDP can be redesigned to be the fastest for CNN and FCL workloads by balancing the count of MAC units to the available memory bandwidth. PIM is not suitable for CNN workloads. PIM is more suitable than CHA for FCL workloads.

\section{Conclusion}

We perform an in-depth analysis of representative state-of-the-art DNN hardware accelerators from Conventional Hardware Accelerator (CHA), Near-Data Processing (NDP), and Processing-in-Memory (PIM) architecture paradigms to identify the fastest and the most energy efficient architecture paradigm for different ML workloads from MLPerf benchmark suite. 
We identified that CHA, on average, has 7$\times$ and 2000$\times$ speedup than NDP and PIM for CNN workloads. In comparison, NDP has, on average 10$\times$ and 40$\times$ speedup than CHA and PIM for workloads with Fully Connected Layers (FCL) such as BERT and DLRM. This inversion is due to the high data reuse potential in CNN workloads, and negligible data reuse potential in FCL, along with the limited bandwidth of the last-level memory.
PIM is the most efficient paradigm and consumes 21$\times$ and 2.7$\times$ lower energy than CHA for CNN and FCL. PIM is also 1.7$\times$ more energy efficient than NDP for all DNN workloads. Further, we identified architectural changes that could be made to these representative architectures to make them faster and more energy-efficient by component-wise sensitivity analysis. In these sensitivity analyses, we have identified conditions for a linear increase in throughput with no loss of latency for FCL workloads for CHAs and a near-linear increase in energy efficiency at no loss of throughput for NDP and PIM paradigms. Further, we have quantitatively established that PIM can not be better than CHA for CNN workloads in terms of latency, while PIM could be better for FCL workloads than CHA.

\bibliographystyle{unsrt}

\end{document}